\documentclass[]{aastex631}

 \hypersetup{linkcolor=red,citecolor=blue,filecolor=cyan,urlcolor=magenta}

\usepackage{amsmath}
\usepackage{soul}

\begin{document}

\title{The effect of plasma-$\beta$ on the heating mechanisms during magnetic reconnection in partially ionized low solar atmosphere}

\author{Abdullah Zafar}
\affiliation{Yunnan Observatories, Chinese Academy of Science,
Kunming, Yunnan 650216 PR china}

\author{Lei Ni}
\affiliation{Yunnan Observatories, Chinese Academy of Science,
Kunming, Yunnan 650216 PR china} 
\affiliation{Yunnan Key Laboratory of Solar Physics and Space Science, Kunming 650216, PR China}
\affiliation{University of Chinese Academy of Sciences, Beijing 100049, PR China}
\correspondingauthor{Lei Ni}
\email{leini@ynao.ac.cn}

\author{Kaifeng Kang}
\affiliation{Yunnan Observatories, Chinese Academy of Science,
Kunming, Yunnan 650216 PR china}
\affiliation{Yunnan Key Laboratory of Solar Physics and Space Science, Kunming 650216, PR China}

\author{Guanchong Cheng}
\affiliation{Yunnan Observatories, Chinese Academy of Science,
Kunming, Yunnan 650216 PR china}
\affiliation{Yunnan Key Laboratory of Solar Physics and Space Science, Kunming 650216, PR China}

\author{Jing Ye}
\affiliation{Yunnan Observatories, Chinese Academy of Science,
Kunming, Yunnan 650216 PR china}
\affiliation{Yunnan Key Laboratory of Solar Physics and Space Science, Kunming 650216, PR China}
\affiliation{University of Chinese Academy of Sciences, Beijing 100049, PR China}

\author{Jun Lin}
\affiliation{Yunnan Observatories, Chinese Academy of Science,
Kunming, Yunnan 650216 PR china}
\affiliation{Yunnan Key Laboratory of Solar Physics and Space Science, Kunming 650216, PR China}
\affiliation{University of Chinese Academy of Sciences, Beijing 100049, PR China}

\author{Ahmad Ali}
\affiliation{Pakistan Tokamak Plasma Research Institute, Islamabad 3329, Pakistan}

\author{Nadia Imtiaz}
\affiliation{Theoretical Physics Division, PINSTECH, Islamabad, Pakistan}

\begin{abstract}

We performed numerical simulations of magnetic reconnection with different strength of magnetic fields from the solar photosphere to the upper chromosphere.
The main emphasis is to identify dominant mechanisms for heating plasmas in the reconnection region under different plasma-$\beta$ conditions in the partially ionized low solar atmosphere.
The numerical results show that more plasmoids are generated in a lower $\beta$ reconnection event. The frequent coalescence of these plasmoids leads to a significant enhancement of turbulence and compression heating, which becomes the dominant mechanism for heating plasma in a lower plasma-$\beta$ reconnection process.
The average power density of the compression heating (Q$_{comp}$) decreases with increasing initial plasma-$\beta$ as a power function: Q$_{comp} \sim \beta_{0}^{-a}$, where the value $a$ is $1.9$ in the photosphere and decreases to about 1.29 in the upper chromosphere. In the photosphere and lower chromosphere, the joule heating contributed by electron-neutral collisions Q$_{en}=\eta_{en} J^2$ eventually dominates over the compression heating when the initial plasma-$\beta$ is larger than the critical value $\beta_{0-critical} = 8$. In the upper chromosphere, the ambipolar diffusion heating and the viscous heating will become equally important as the compression heating when the initial plasma-$\beta$ is larger than the critical value $\beta_{0-critical} = 0.5$. These results indicate that the compression heating caused by turbulent reconnection mediated with plasmoids is likely the major heating mechanism for the small-scale reconnection events with stronger magnetic fields such as active region EBs and UV bursts.  However, the heating caused by the partial ionization effects can not be ignored for those reconnection events with weaker magnetic fields such as quiet Sun EBs and cold surges.

\end{abstract}


\section{Introduction} \label{sec:intro}
Magnetic reconnection, a fundamental process in magnetized plasmas where the merging of oppositely directed magnetic field lines leads to the rearrangement of the magnetic field configuration and the conversion of stored magnetic energy into kinetic and thermal energy of the plasma~\citep{priest2007magnetic}.
Magnetic reconnection occurs in various environments, spanning from laboratory to astrophysical plasmas, and has been extensively studied in the fully ionized plasma in the past 50 years~\citep{zweibel2009magnetic,katz2010laboratory,cassak2012magnetic}.
However, research on magnetic reconnection in the cooler plasma environments with partially ionized plasmas has received less attention, with relatively few studies focused on it.
Investigating magnetic reconnection in partially ionized plasmas is more challenging because of the additional interactions between charged particles and neutrals, which can influence the dynamics of the reconnection process.
Thus, how neutrals affect the reconnection process is one of the ten major challenges in the field of magnetic reconnection~\citep{2020arXiv200908779J}.
Partially ionized plasmas are found in a wide range of astrophysical conditions, such as the lower solar atmosphere, protoplanetary nebulae, comet tails, young star objects, and the interstellar medium (ISM)~\citep{ballester2018partially}.
This motivates researchers to study magnetic reconnection in partially ionized plasmas to gain a better understanding of the energy release processes in these cooler plasma environments.

The plasma characteristics in the cool and dense lower solar atmosphere, including the photosphere and chromosphere, differ significantly compared to those in the much hotter and rarefied corona.
This thin layer, roughly 2000 km above the Sun surface, is filled with highly density stratified and partially ionized plasmas.
The hydrogen density decreases drastically with height, and its ionization degree normally varies from $10^{-4}$ at around the solar temperature minimum region to about $1$ in the upper chromosphere~\citep{ni2020magnetic}.
Numerous small-scale events associated with magnetic reconnection have been observed in the highly dynamic lower solar atmosphere to date.
These include Ellerman-bombs (EBs)~\citep{ellerman1917solar,georgoulis2002statistics}, quiet Sun Ellerman bombs (QSEBs)~\citep{van2016reconnection,nelson2017iris,joshi2020signatures}, ultraviolet (UV) bursts~\citep{peter2014hot,tian2016iris}, chromospheric jets~\citep{shibata2007chromospheric,singh2011chromospheric}, and some other transient events~\citep{xue2021decay,rast2021critical}.
The temperature increase and dynamic process in these small-scale magnetic reconnection events are strongly influenced by plasma-$\beta$, which is a function of the magnetic field strength, plasma density, and temperature~\citep{ni2012effects,peter2019plasmoid,liu2023numerical,danilovic2017simulating}.
The previous single and multi-fluids MHD simulations showed that the plasmas can be heated to a temperature above 20,000 K in the lower chromosphere with high plasma density~\citep{ni2016heating,ni2018magnetic,ni2022plausibility}, as long as the strength of the reconnection magnetic field is greater than 500 G.

The fast reconnection and heating mechanisms can also be considerably influenced by many other factors, except plasma-$\beta$. 
For the same plasma-$\beta$, the temperature increase in a reconnection process in the dense photosphere and lower chromosphere is much smaller than that in the upper chromosphere~\citep{zafar2023high}. 
The previous numerical results showed that the magnetic diffusion contributed by electron-neutral collisions can significantly accelerate the reconnection process and cause more plasma heating in the weak reconnection magnetic field case below the middle chromosphere~\citep{liu2023numerical,zafar2023high}. 
Ambipolar diffusion (decoupling of ions and neutrals) can result in much faster thinning of the current sheet in the absence of the guide field, but even a weak guide field can impede its effect~\citep{brandenburg1994formation,heitsch2003fast,heitsch2003suppression,ni2015fast}. 
Recent numerical results demonstrated that the heating contributed by ambipolar diffusion is significant in the case with higher plasma-$\beta$ and guide fields in the upper chromosphere, but it will slow down the reconnection process at the same time~\citep{zafar2023high}. 
The two-fluid simulations showed that ion-recombination in the reconnection site, combined with Alfvenic outflows, rapidly removes ions from the reconnection region, resulting in a fast reconnection rate independent of the Lundquist number~\citep{leake2012multi,leake2013magnetic}. 
The recent advanced multifluid multispecies (helium-hydrogen mixture) simulation results demonstrated that the presence of helium species leads to more efficient heating mechanisms than the two-fluid case in the chromospheric magnetic reconnection, enabling transition-region temperatures and helium enrichment in outflows, potentially explaining solar wind observations~\citep{wargnier2023multifluid,wargnier2025time}.
However, all these partial ionization effects will become less important when the reconnection site is heated to a high temperature, in the case with a strong magnetic field and a lower plasma-$\beta$. 

In this work, we have performed a large number of high-resolution simulation tests which cover the reconnection processes with different plasma-$\beta$ and at different altitudes. 
The partial ionization effects such as magnetic diffusion caused by electron-neutral collision, ambipolar diffusion, and viscosity are all included.
We aim to analyze the influence of the initial plasma-$\beta$ (the strength of the initial magnetic field, b$_0$) on the different heating mechanisms in the reconnection process.
Moreover, identifying the main plasma heating mechanisms in the reconnection process at different low solar atmosphere altitudes under various plasma-$\beta$ conditions is also the focus of attention.
The numerical model and initial setup is described in Section~\ref{sec_II}.
The numerical results are presented in Sections~\ref{sec_III}.
Finally, a summary and conclusion of the study is given in Section~\ref{sec_IV}.

\section{Numerical model and initial setup}
\label{sec_II}
High-resolution 2.5D numerical experiments are carried out with the single-fluid MHD code, NIRVANA~\citep{ziegler2011semi}.
It is presumed that all the species of hydrogen-helium plasma ($H$, $H^{+}$, $H_{e}$, $H_{e}^{+}$ and electrons) are well coupled and therefore treated as a single fluid.
The following single-fluid MHD equations are used here:
\begin{eqnarray}
\frac{\partial \rho}{\partial t} = - \nabla \cdot (\rho \mathbf{v}),
\label{eq:1}
\end{eqnarray}

\begin{eqnarray}
\begin{split}   
\frac{\partial (\rho \mathbf{v})}{\partial t} & = - \nabla \cdot \left[\rho \mathbf{vv}+\left(p+\frac{1}{2\mu_{0}} |\mathbf{B}|^{2}\right)I-\frac{1}{\mu_{0}}\mathbf{BB}\right] \\
&+\nabla \cdot \tau_{S},
\end{split}
\label{eq:2}
\end{eqnarray}

\begin{eqnarray}
\begin{split}
    \frac{\partial e}{\partial t} &= -\nabla \cdot \left[\left(e+p+\frac{1}{2\mu_{0}} |\mathbf{B}|^{2}\right) \mathbf{v}\right] \\
    & + \nabla \cdot \left[\frac{1}{\mu_{0}} (\mathbf{v} \cdot \mathbf{B})\mathbf{B}\right] \\
    & + \nabla \cdot \left[\mathbf{v} \cdot \tau_{s} + \frac{\eta}{\mu_{0}} \mathbf{B} \times (\nabla \times \mathbf{B})\right] \\
    & - \nabla \cdot \left[\frac{1}{\mu_{0}} \mathbf{B} \times \mathbf{E}_{AD}\right] \\
    & + Q_{rad} +\mathcal{H}, 
\end{split}
\label{eq:3}
\end{eqnarray}

\begin{eqnarray}
\frac{\partial \mathbf{B}}{\partial t} = \nabla \times (\mathbf{v} \times \mathbf{B} - \eta \nabla \times \mathbf{B} + \mathbf{E}_{AD}),
\label{eq:4}
\end{eqnarray}
with
\begin{eqnarray}
e = \frac{p}{\gamma -1} + \frac{1}{2} \rho |\mathbf{v}|^{2} + \frac{1}{2 \mu_{0}} |\mathbf{B}|^{2}
\label{eq:5}
\end{eqnarray}
and
\begin{eqnarray}
p = \frac{(1.1+Y_{iH}+0.1Y_{iHe})\rho}{1.4 m_{i}} k_{B} T,
\label{eq:6}
\end{eqnarray}
where $\rho$ is the mass density, $\mathbf{v}$ is the plasma velocity, $\mathbf{B}$ is the magnetic field, $p$ is the thermal pressure, $e$ represents the total energy density, $T$ is the plasma temperature; and $Y_{iH}$ and $Y_{iHe}$ denote the ionization fractions of hydrogen and helium, respectively, whereas m$_{i}$ is the proton mass, and $k_{B}$ is the Boltzmann constant.
The total number density of helium is set to 10\% compared to hydrogen, and only primary helium ionization is taken into account.
The adiabatic constant $\gamma$ is set to 5/3. The stress tensor, $\tau_{S} = \xi [\nabla \mathbf{v} + (\nabla \mathbf{v})^{T} - \frac{2}{3} (\nabla \cdot \mathbf{v})I]$ , where $\xi$ is the coefficient of dynamic viscosity, having units of kg m$^{-1}$ s$^{-1}$.
The current sheet investigated in the present work is parallel to the Sun surface, and its width reduces to less than several tens of kilometers in the main reconnection process; therefore, the gravity effect is excluded, and the initial plasma temperature and density are supposed to be uniform throughout the simulation domain.

The interactions between different species in partially ionized plasma (ions, electrons, and neutrals) are of significant importance to understand the dynamics of the magnetic reconnection process in the low solar atmosphere.
The total diffusion caused by the collision of electrons with ions ($\eta_{ei}$) and neutrals ($\eta_{en}$) is known as magnetic diffusion ($\eta$). It is expressed as~\citep{khomenko2012heating}
\begin{eqnarray}
\eta = \eta_{ei} + \eta_{en} = \frac{m_{e} \nu_{ei}}{e^{2}_{c} n_{e} \mu_{0}} + \frac{m_{e} \nu_{en}}{e^{2}_{c} n_{e} \mu_{0}}
\label{eq:7}
\end{eqnarray}
where $m_{e}$, $e_{c}$, $\mu_{0}$, $n_{e}$, $\nu_{en}$, and $\nu_{ei}$ are the mass of the electron, the electron charge, the permeability of free space, the electron density and the collision frequency of the electrons with neutrals and ions, respectively.
The electron number density, electron-ion, and electron-neutral collision frequencies are~\citep{braginskii1965transport,spitzer1962jr,hunana2022generalized}

\begin{eqnarray}
n_{e} = \frac{\rho(Y_{iH}+0.1Y_{iHe})}{1.4m_{i}}, 
\label{eq:8}
\end{eqnarray}

\begin{eqnarray}
 \nu_{ei} = \frac{n_{e} e^{4}_{c} \Lambda}{3 m^{2}_{e} \epsilon^{2}_{0}} \left( \frac{m_{e}}{2 \pi k_{B} T} \right)^{3/2}, 
\label{eq:9}
\end{eqnarray}
and
\begin{eqnarray}
\nu_{en} = n_{n} \sqrt{\frac{8 k_{B} T}{\pi m_{e}}} \sigma_{en}.
\label{eq:10}
\end{eqnarray}
where $\Lambda$, $\epsilon_{0}$, $n_{n}$, and $\sigma_{en}$ represent the Coulomb logarithm, permittivity of free space, number density of neutral species, and electron-neutral collision cross-section, respectively. The expression for $\Lambda$ is~\citep{khomenko2012heating}
\begin{eqnarray}
\Lambda = 23.4-1.15 \log_{10} n_{e} + 3.45 \log_{10} T.
\label{eq:11}
\end{eqnarray}

In this study, we are dealing with a hydrogen and helium plasma mixture, therefore the electron-neutral collision frequency ($\nu_{en}$) comes from the collision of an electron with neutral hydrogen as well as neutral helium and is

\begin{eqnarray}
\nu_{en} = n_{n H} \sqrt{\frac{8 k_{B} T}{\pi m_{e}}} \sigma_{e-n H}+ n_{n H_{e}} \sqrt{\frac{8 k_{B} T}{\pi m_{e}}} \sigma_{e-n H_{e}},
\label{eq:12}
\end{eqnarray}
where $n_{nH} = \rho(1-Y_{iH})/(1.4mi)$ and $n_{nHe} = 0.1\rho(1-Y_{iHe})/(1.4mi)$ are the neutral hydrogen and helium number densities, respectively.
Fixed collision cross-sections for electron-neutral hydrogen ($\sigma_{e-n H}$) and electron-neutral helium ($\sigma_{e-n H_{e}}$) are employed here, and their values are $2 \times 10^{-19}$ m$^{2}$ and $\sigma_{e-n H}/3$, respectively~\citep{vranjes2013collisions}.
Such simplified cross-section models are different from those in the recent work~\citep{wargnier2022detailed}, where they used the temperature-dependent collision cross-sections ($\sigma_{e-n H}$ and $\sigma_{e-n H_{e}}$). The magnetic diffusivities are then derived as: 
\begin{eqnarray}
\eta_{ei} \simeq 1.0246 \times 10^{8} \Lambda T^{-1.5},
\label{eq:13}
\end{eqnarray}
and
\begin{eqnarray}
\eta_{en} \simeq 0.0351 \sqrt{T} \frac{\left[\frac{0.1}{3} (1-Y_{iH_{e}})+(1-Y_{iH})\right]}{Y_{iH} + 0.1 Y_{iHe}},
\label{eq:14}
\end{eqnarray}
where $\eta_{ei}$ and $\eta_{en}$ have dimensions of m$^{2}$ s$^{-1}$.

The energy and induction equations~(Eqs.\ref{eq:3} and \ref{eq:4}) both contain the ambipolar electric field (E$_{AD}$)~\citep{braginskii1965transport}
\begin{eqnarray}
\mathbf{E}_{AD} = \frac{1}{\mu_{0}} \eta_{AD} [(\nabla \times B) \times B ] \times B,
\label{eq:15}
\end{eqnarray}
here $\eta_{AD}$ denotes the ambipolar diffusion coefficient and is written as~\citep{khomenko2012heating,hunana2022generalized}
\begin{eqnarray}
\eta_{AD} = \frac{(\rho_{n}/\rho)^2}{\rho_{i} \nu_{in} + \rho_{e} \nu_{en}}
\label{eq:16}
\end{eqnarray}
in units of m$^{3}$ s kg$^{-1}$. For plasma containing both helium and hydrogen, $\rho_{n}/\rho$ equals~\citep{ni2022plausibility}
\begin{eqnarray}
\rho_{n}/\rho  = \frac{0.4 (1- Y_{iHe}) + (1- Y_{iH})}{1.4}.
\label{eq:17}
\end{eqnarray}
The ion collision part is expressed as~\citep{spitzer1962jr}
\begin{eqnarray}
\begin{split}
    \rho_{i} \nu_{in} & = \rho_{iH} n_{nH} \sqrt{\frac{8 k_{B} T}{\pi m_{i}/2}} \sigma_{iH-nH} \\
    & + \rho_{iH} n_{nHe} \sqrt{\frac{8 k_{B} T}{4 \pi m_{i}/5}} \sigma_{iH-nHe} \\
    & + \rho_{iHe} n_{nH} \sqrt{\frac{8 k_{B} T}{4 \pi m_{i}/5}} \sigma_{iHe-nH} \\
    & + \rho_{iHe} n_{nHe} \sqrt{\frac{8 k_{B} T}{2 \pi m_{i}}} \sigma_{iHe-nHe}, 
\end{split}
\label{eq:18}
\end{eqnarray}
where $\rho_{iH} = \rho Y_{iH}/1.4$ represents ionized hydrogen mass density and $\rho_{iHe} = 0.4\rho Y_{iHe}/1.4$ is the ionized helium mass density.
The other parameters $\sigma_{iH-nH}$ and $\sigma_{iH-nHe}$ denote the collision cross-section of ionized hydrogen with neutral hydrogen and helium, respectively; whereas $\sigma_{iHe-nH}$ and $\sigma_{iHe-nHe}$ are the collision cross-section of ionized helium with neutral hydrogen and helium.
Their values are $\sigma_{iH-nH} = 1.5 \times 10^{-18}$ m$^{2}$, $\sigma_{iH-nHe}$ = $\sigma_{iHe-nH}$ = $\sigma_{iHe-nHe}$ = $\sigma_{iH-nH}/\sqrt{3}$~\citep{vranjes2013collisions,barata2010elastic}.
The electron collision part is~\citep{spitzer1962jr,ni2022plausibility}
\begin{eqnarray}
\begin{split}
    \rho_{e} \nu_{en} & = \rho_{e} n_{nH} \sqrt{\frac{8 k_{B} T}{\pi m_{e}}} \sigma_{e-nH} \\
    & + \rho_{e} n_{nHe} \sqrt{\frac{8 k_{B} T}{\pi m_{e}}} \sigma_{e-nHe}, \\
\end{split}
\label{eq:19}
\end{eqnarray}
where $\sigma_{e-nH}$ is the $e-nH$ collision cross-section and $\sigma_{e-nHe}$ refers to the $e-nHe$ collision cross-sections.
These collision cross-sections ($\sigma_{e-nH}$, $\sigma_{e-nHe}$) are smaller compared to $\sigma_{iH-nH}$ and hence are neglected. The ambipolar diffusion coefficient simplified into
\begin{eqnarray}
\begin{split}
    \eta_{AD} = \frac{(\rho_{n}/\rho)^2}{\rho_{i} \nu_{in}}.
\end{split}
\label{eq:19a}
\end{eqnarray}

When dealing with partially ionized plasmas, both ion and neutral species are contributing to the dynamic viscosity, which is~\citep{leake2013magnetic}
\begin{eqnarray}
\xi = \xi_{i} + \xi_{n} = \frac{n_{i} k_{B} T}{\nu_{ii}} + \frac{n_{n} k_{B} T}{\nu_{nn}},
\label{eq:20}
\end{eqnarray}
where $\xi_{i,n}$ is the viscosity coefficient of the ions and neutrals, and $\nu_{ii,nn}$ represents the collision frequencies between the ion-ion and the neutral-neutral.
The collision frequencies between these similar plasma species are expressed as~\citep{leake2013magnetic}:
\begin{eqnarray}
\nu_{nn} = n_{n} \sigma_{nn} \sqrt{\frac{16 k_{B} T}{\pi m_{n}}}
\label{eq:21}
\end{eqnarray}
and
\begin{eqnarray}
\nu_{ii} = \frac{n_{i} e^{4}_{c} \Lambda}{3 m^{2}_{i} \epsilon^{2}_{0}} \left(\frac{m_{i}}{2 \pi k_{B} T} \right)^{3/2}.
\label{eq:22}
\end{eqnarray}
The final form of the dynamic viscosity used in our numerical calculations is~\citep{zafar2023high,zafar2024unraveling}
\begin{eqnarray}
\xi = \xi_{i} + \xi_{n} \simeq  \frac{4.8692 \times 10^{-16}}{\Lambda} T^{2} \sqrt{T}+ 2.0127 \times 10^{-7} \sqrt{T}. 
\label{eq:24}
\end{eqnarray}

This study uses temperature-dependent hydrogen and helium ionization degrees.
The modified Saha and Boltzamann equation~\citep{gan1990hydrodynamic} is used to determine the ionization degree for hydrogen in the photosphere.
Whereas, for the chromospheric plasma, the hydrogen ionization degree is obtained from the table provided by~\cite{carlsson2012approximations}.
The temperature-dependent helium ionization degree is $Y_{iHe} = 1-10^{0.325571-0.0000596T}$.
We set $Y_{iHe} = 0.00010084814$ for the plasma temperature less then 5413 K.
At the start of the simulations, ionization degrees across the simulation domain are constant because of the uniform initial plasma density and temperature profiles; however, these ionization degrees are later modified according to plasma local parameters.

Radiative transfer processes between the radiation field and the solar atmosphere are significant in the lower solar atmosphere.
The Gan \& Fang model~\citep{gan1990hydrodynamic} and the Carlsson \& Leenaarts model~\citep{carlsson2012approximations}, are included in our photospheric and chromospheric simulations, respectively.
Details about the implementation of these cooling models are provided in our previous studies~\citep{zafar2023high,zafar2024unraveling}.

Numerical simulations are conducted using different initial plasma-$\beta$ at multiple photospheric and chromospheric altitudes to examine its influence on the evolution of various heating mechanisms in the reconnection process.
All simulations started with an antiparallel magnetic field topology of the force-free Harris sheet, expressed as $B_{x0} = -b_{0} \tanh [y/(0.05L_{0})]$, $B_{y0} = 0$, and $B_{z0} = b_{0}/ \cosh [y/(0.05L_{0})]$, where b$_0$ represents the initial magnetic field strength and L$_{0} = 2 \times 10^5 m$ is the system scale length.
The initial plasma-$\beta$ varies by changing initial magnetic field (Table~\ref{tab:table1}).
The simulation box used here ranges from 0 to $L_{0}$ along the $x$-axis and from $-0.5L_{0}$ to $0.5L_{0}$ in the $y$-axis, comparable to the length scale of small-scale magnetic reconnection activities observed in the lower solar atmosphere.
Open boundary conditions are employed along both axes.
Initially, a small magnetic perturbation is introduced to initiate the reconnection process, given as $b_{x1} = - b_{pert} \sin \left[{2 \pi (y+0.5L_{0})}/{L_{0}}\right] \cos \left[{2 \pi (x+0.5L_{0})}/{L_{0}}\right]$ and $b_{y1} = b_{pert} \cos \left[{2 \pi (y+0.5L_{0})}/{L_{0}}\right] \sin \left[{2 \pi (x+0.5L_{0})}/{L_{0}}\right]$ with b$_{pert}$  = 0.005 b$_{0}$.
A base-level grid of 192 $\times$ 192 with an adaptive mesh refinement (AMR) level of 9 is utilized here.
The initial temperature (T$_{0}$) and mass density ($\rho_{0}$) at each altitude in the lower solar atmosphere, as shown in Table~\ref{tab:table1}, are obtained from the C7 atmosphere model~\citep{avrett2008models}.

\begin{table}
\caption{\label{tab:table1} Initial values of mass density ($\rho_{0}$), temperature (T$_{0}$), ionization fraction of hydrogen (Y$_{iH0}$), ionization fraction of helium (Y$_{iHe0}$), collision frequencies of ionized hydrogen and neutral hydrogen ($\nu_{iH-nH0}$), ionized hydrogen and neutral helium ($\nu_{iH-nHe0}$), ionized helium and neutral hydrogen ($\nu_{iHe-nH0}$), ionized helium and neutral helium ($\nu_{iHe-nHe0}$), and magnetic field (b$_{0}$) used in the simulations.}
  \centerline{%
  \resizebox{\textwidth}{!}{
    \begin{tabular}{|l|*{6}{c|}} 
      \hline
    \cline{2 - 5}
    \hline
  \begin{tabular}[c]{@{}c@{}}Z (km)\end{tabular} & 400 & 600 & 800 & 1000 & 1400 & 2000 \\
  \hline
  \begin{tabular}[c]{@{}c@{}}$\rho_{0}$ (kg m$^{-3}$)\end{tabular} & 1.56 $\times$ 10$^{-5}$ & 2.40 $\times$ 10$^{-6}$ & 3.44 $\times$ 10$^{-7}$ & 6.32 $\times$ 10$^{-8}$ & 4.51 $\times$ 10$^{-9}$ & 1.68 $\times$ 10$^{-10}$ \\
  \hline
  \begin{tabular}[c]{@{}c@{}}T$_0$ (K)\end{tabular} & 4590 & 4421 & 5100 & 6223 & 6610 & 6678 \\
  \hline
  \begin{tabular}[c]{@{}c@{}}Y$_{iH0}$\end{tabular} & 2.532 $\times$ 10$^{-5}$ & 3.912 $\times$ 10$^{-5}$ & 2.500 $\times$ 10$^{-5}$ & 1.183 $\times$ 10$^{-2}$ & 6.242 $\times$ 10$^{-2}$ & 7.453 $\times$ 10$^{-2}$ \\
  \hline
  \begin{tabular}[c]{@{}c@{}}Y$_{iHe0}$\end{tabular} & 1.00848 $\times$ 10$^{-4}$ & 1.00848 $\times$ 10$^{-4}$ & 1.00848 $\times$ 10$^{-4}$ & 1.0529 $\times$ 10$^{-1}$ & 1.516 $\times$ 10$^{-1}$ & 1.5945 $\times$ 10$^{-1}$ \\
  \hline 
  \begin{tabular}[c]{@{}c@{}}$\nu_{iH-nH0}$ (s$^{-1}$)\end{tabular} & 1.407 $\times$ 10$^{8}$ & 2.119 $\times$ 10$^{7}$ & 3.269 $\times$ 10$^{6}$ & 6.542 $\times$ 10$^{5}$ & 4.566 $\times$ 10$^{4}$ & 1.686 $\times$ 10$^{3}$ \\
  \hline 
  \begin{tabular}[c]{@{}c@{}}$\nu_{iH-nHe0}$ (s$^{-1}$)\end{tabular} & 6.422 $\times$ 10$^{6}$ & 9.674 $\times$ 10$^{5}$ & 1.492 $\times$ 10$^{5}$ & 2.703 $\times$ 10$^{4}$ & 1.886 $\times$ 10$^{3}$ & 6.989 $\times$ 10$^{1}$ \\
  \hline
  \begin{tabular}[c]{@{}c@{}}$\nu_{iHe-nH0}$ (s$^{-1}$)\end{tabular} & 6.422 $\times$ 10$^{7}$ & 9.675 $\times$ 10$^{6}$ & 1.492 $\times$ 10$^{6}$ & 2.986 $\times$ 10$^{5}$ & 2.084 $\times$ 10$^{4}$ & 7.695 $\times$ 10$^{2}$ \\
  \hline  
  \begin{tabular}[c]{@{}c@{}}$\nu_{iHe-nHe0}$ (s$^{-1}$)\end{tabular} & 4.062 $\times$ 10$^{6}$ & 6.118 $\times$ 10$^{5}$ & 9.437 $\times$ 10$^{4}$ & 1.709 $\times$ 10$^{4}$ & 1.193 $\times$ 10$^{3}$ & 4.420 $\times$ 10$^{1}$ \\
  \hline   
  \begin{tabular}[c]{@{}c@{}}$\beta_0 = 0.01$ \\ b$_0$ (T)\end{tabular} & -- & -- & -- & -- & -- & 1.4 $\times$ 10$^{-3}$ \\
  \hline
  \begin{tabular}[c]{@{}c@{}}$\beta_0 = 0.05$ \\ b$_0$  (T)\end{tabular} & 1.54 $\times$ 10$^{-1}$ & 5.9 $\times$ 10$^{-2}$ & 2.4 $\times$ 10$^{-2}$ & 1.2 $\times$ 10$^{-2}$ & 3.2 $\times$ 10$^{-3}$ & 6.3 $\times$ 10$^{-4}$ \\
  \hline
  \begin{tabular}[c]{@{}c@{}}$\beta_0 = 0.5$ \\ b$_0$ (T)\end{tabular} & 4.9 $\times$ 10$^{-2}$ & 1.9 $\times$ 10$^{-2}$ & 7.6 $\times$ 10$^{-3}$ & 3.6 $\times$ 10$^{-3}$ & 1.0 $\times$ 10$^{-3}$ & 2.0 $\times$ 10$^{-4}$ \\
  \hline
  \begin{tabular}[c]{@{}c@{}}$\beta_0 = 2.0$ \\ b$_0$ (T)\end{tabular} & 2.5 $\times$ 10$^{-2}$ & 9.3 $\times$ 10$^{-3}$ & 3.8 $\times$ 10$^{-3}$ & 1.8 $\times$ 10$^{-3}$ & 5.1 $\times$ 10$^{-4}$ & 1.0 $\times$ 10$^{-4}$ \\
  \hline
  \begin{tabular}[c]{@{}c@{}}$\beta_0 = 3.0$ \\ b$_0$ (T)\end{tabular} & 2.0 $\times$ 10$^{-2}$ & 7.6 $\times$ 10$^{-3}$ & 3.1 $\times$ 10$^{-3}$ & 1.5 $\times$ 10$^{-3}$ & 4.2 $\times$ 10$^{-4}$ & 8.1 $\times$ 10$^{-5}$ \\
  \hline
  \begin{tabular}[c]{@{}c@{}}$\beta_0 = 4.0$ \\ b$_0$ (T)\end{tabular} & 1.7 $\times$ 10$^{-2}$ & 6.6 $\times$ 10$^{-3}$ & -- & -- & -- & -- \\
  \hline
  \begin{tabular}[c]{@{}c@{}}$\beta_0 = 5.0$ \\ b$_0$ (T)\end{tabular} & 1.5 $\times$ 10$^{-2}$ & 5.9 $\times$ 10$^{-3}$ & 2.4 $\times$ 10$^{-3}$ & 1.1 $\times$ 10$^{-3}$ & 3.2 $\times$ 10$^{-4}$ & 6.3 $\times$ 10$^{-5}$ \\
  \hline
  \begin{tabular}[c]{@{}c@{}}$\beta_0 = 8.0$ \\ b$_0$ (T)\end{tabular} & 1.2 $\times$ 10$^{-2}$ & 4.7 $\times$ 10$^{-3}$ & 1.9 $\times$ 10$^{-3}$ & -- & -- & -- \\
  \hline
  \begin{tabular}[c]{@{}c@{}}$\beta_0 = 10.0$ \\ b$_0$ (T)\end{tabular} & 1.1 $\times$ 10$^{-2}$ & 4.2 $\times$ 10$^{-3}$ & 1.7 $\times$ 10$^{-3}$ & 8.0 $\times$ 10$^{-4}$ & 2.3 $\times$ 10$^{-4}$ & 4.5 $\times$ 10$^{-5}$ \\
  \hline
  \begin{tabular}[c]{@{}c@{}}$\beta_0 = 15.0$ \\ b$_0$ (T)\end{tabular} & -- & -- & 1.4 $\times$ 10$^{-3}$ & 6.6 $\times$ 10$^{-4}$ & 1.9 $\times$ 10$^{-4}$ & 3.6 $\times$ 10$^{-5}$ \\
  \hline
    \end{tabular}%
    }
   }
   \label{table1}
\end{table}

\begin{figure}
\centering
\begin{minipage}{0.494\textwidth}
\includegraphics[width=1.0\textwidth]{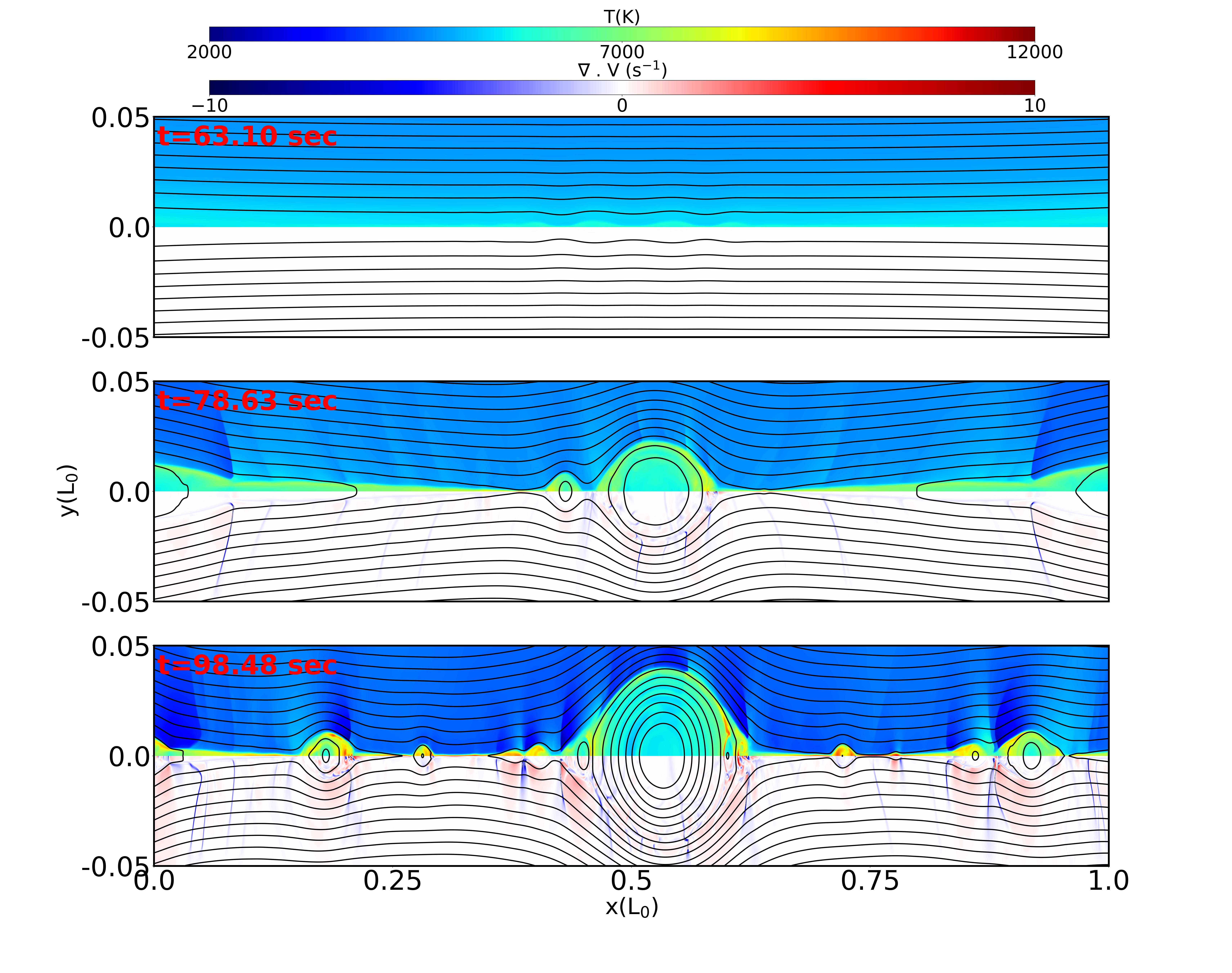}
\put(-185,203){\textbf{(a) Z = 400 km, $\beta_0$ = 0.5}}
\end{minipage}
\begin{minipage}{0.494\textwidth}
\includegraphics[width=1.0\textwidth]{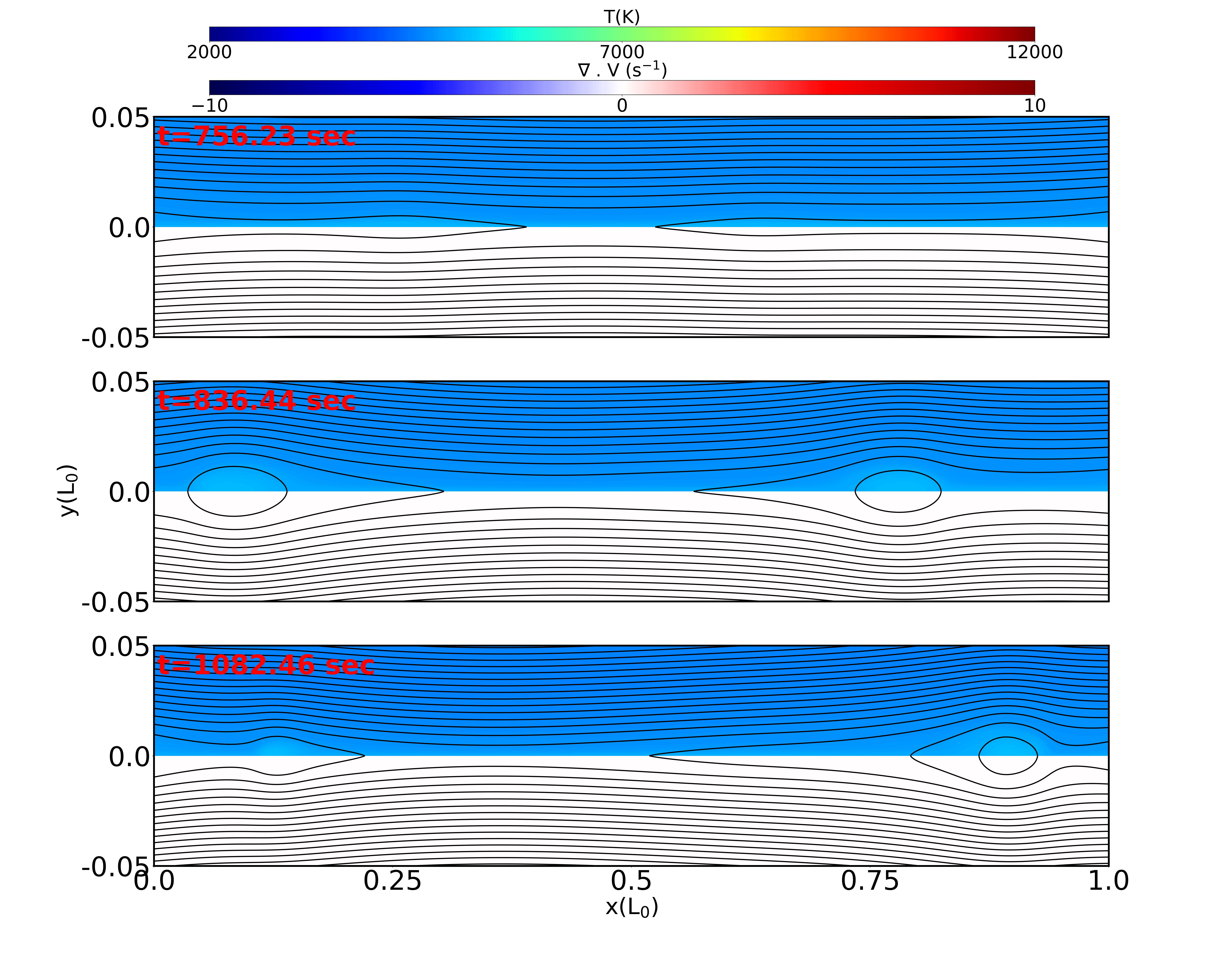}
\put(-185,203){\textbf{(b) Z = 400 km, $\beta_0$ = 10.0}}
\end{minipage}
\caption{Contours of temperature (upper half panel) and $\nabla \cdot$ V (lower half panel) at three different stages during the reconnection process with initial plasma-$\beta$ of 0.5 and 10.0 at the photosphere (Z = 400 km). The black lines illustrate the distribution of magnetic field lines.}
\label{fig_1}
\end{figure}
\begin{figure}
\centering
\begin{minipage}{0.494\textwidth}
\includegraphics[width=1.0\textwidth]{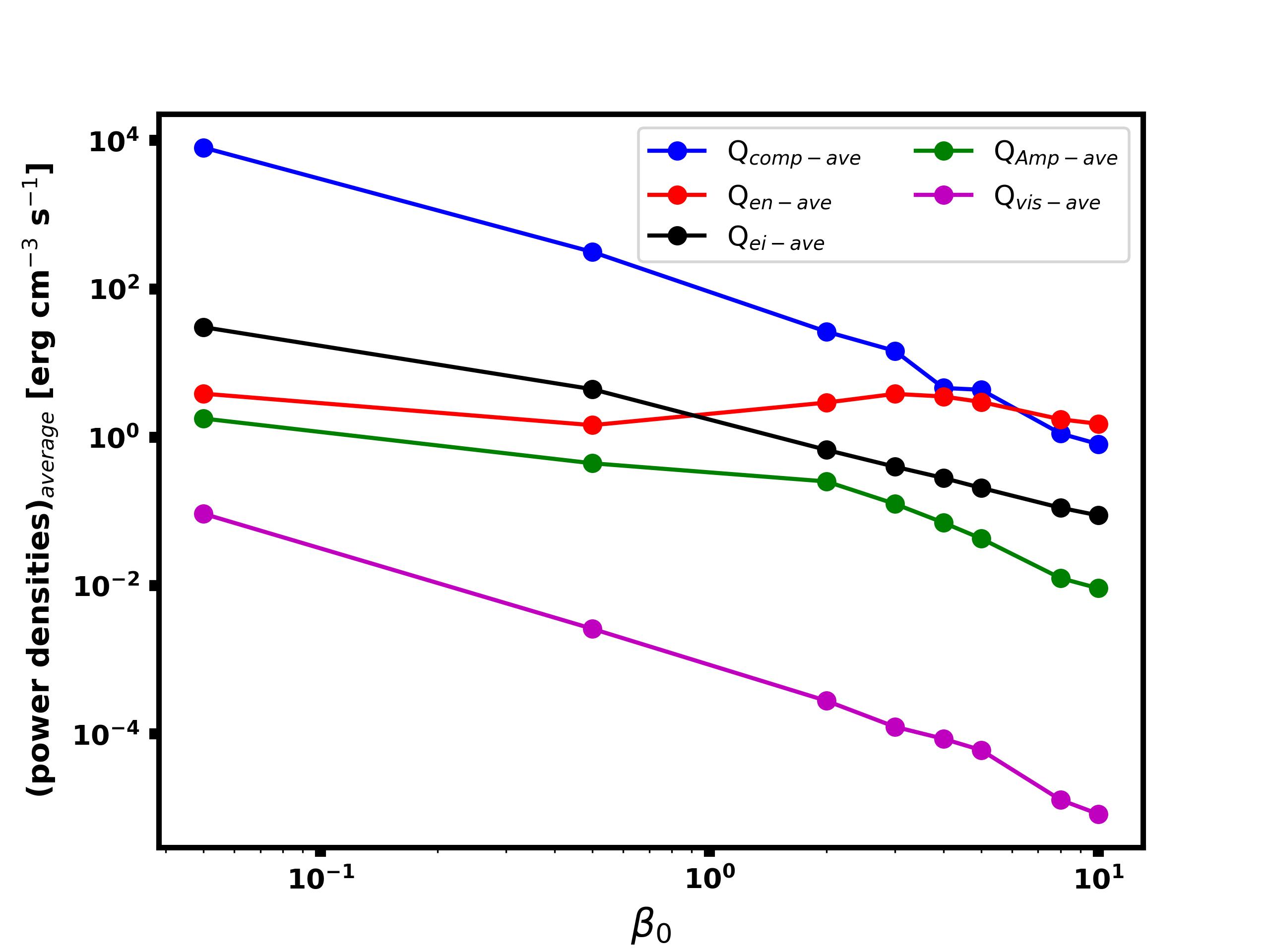}
\put(-160,175){\textbf{(a) Z = 400 km}}
\end{minipage}
\begin{minipage}{0.494\textwidth}
\includegraphics[width=1.0\textwidth]{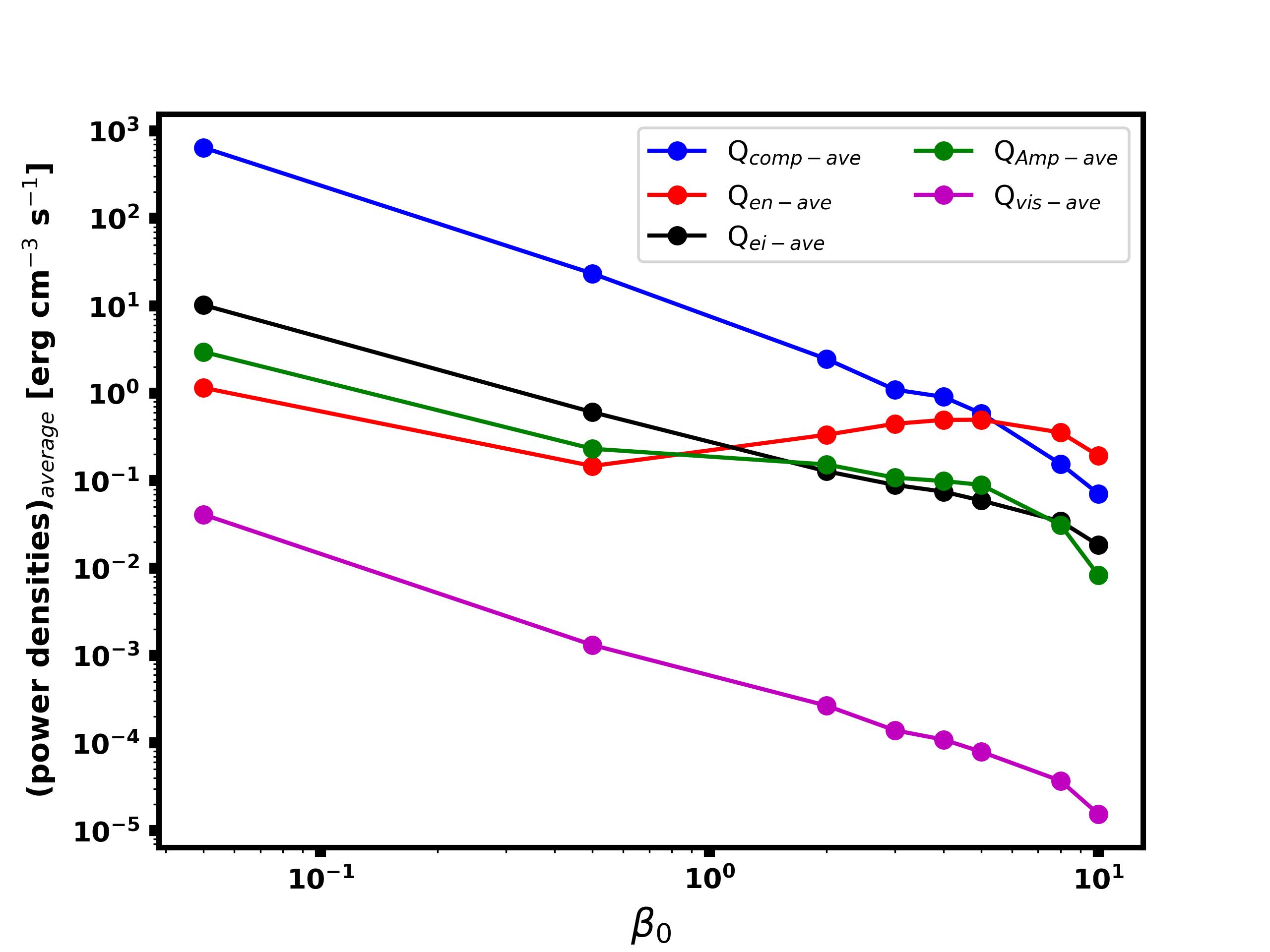}
\put(-160,175){\textbf{(b) Z = 600 km}}
\end{minipage}
\begin{minipage}{0.494\textwidth}
\includegraphics[width=1.0\textwidth]{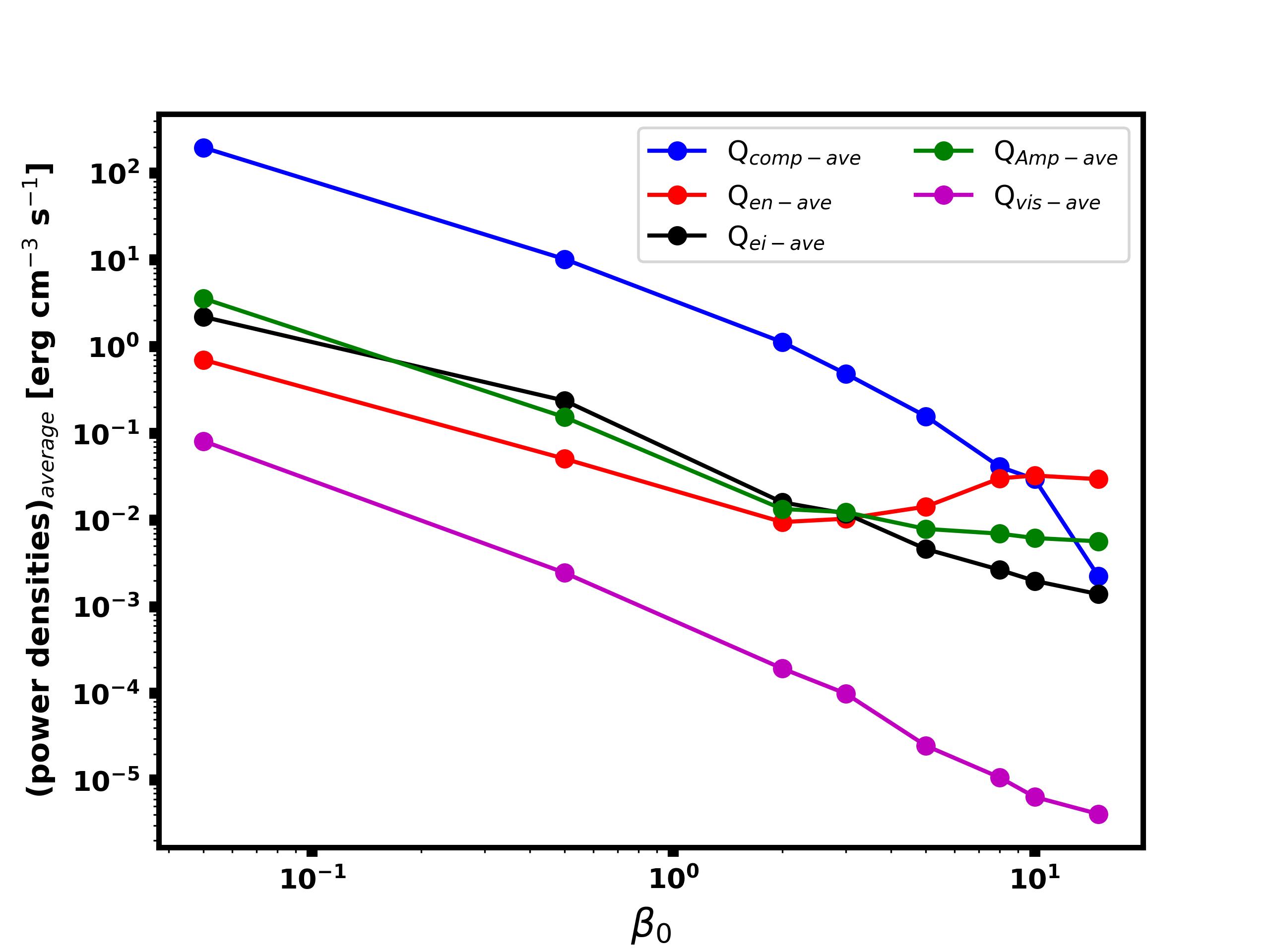}
\put(-160,175){\textbf{(c) Z = 800 km}}
\end{minipage}
\begin{minipage}{0.494\textwidth}
\includegraphics[width=1.0\textwidth]{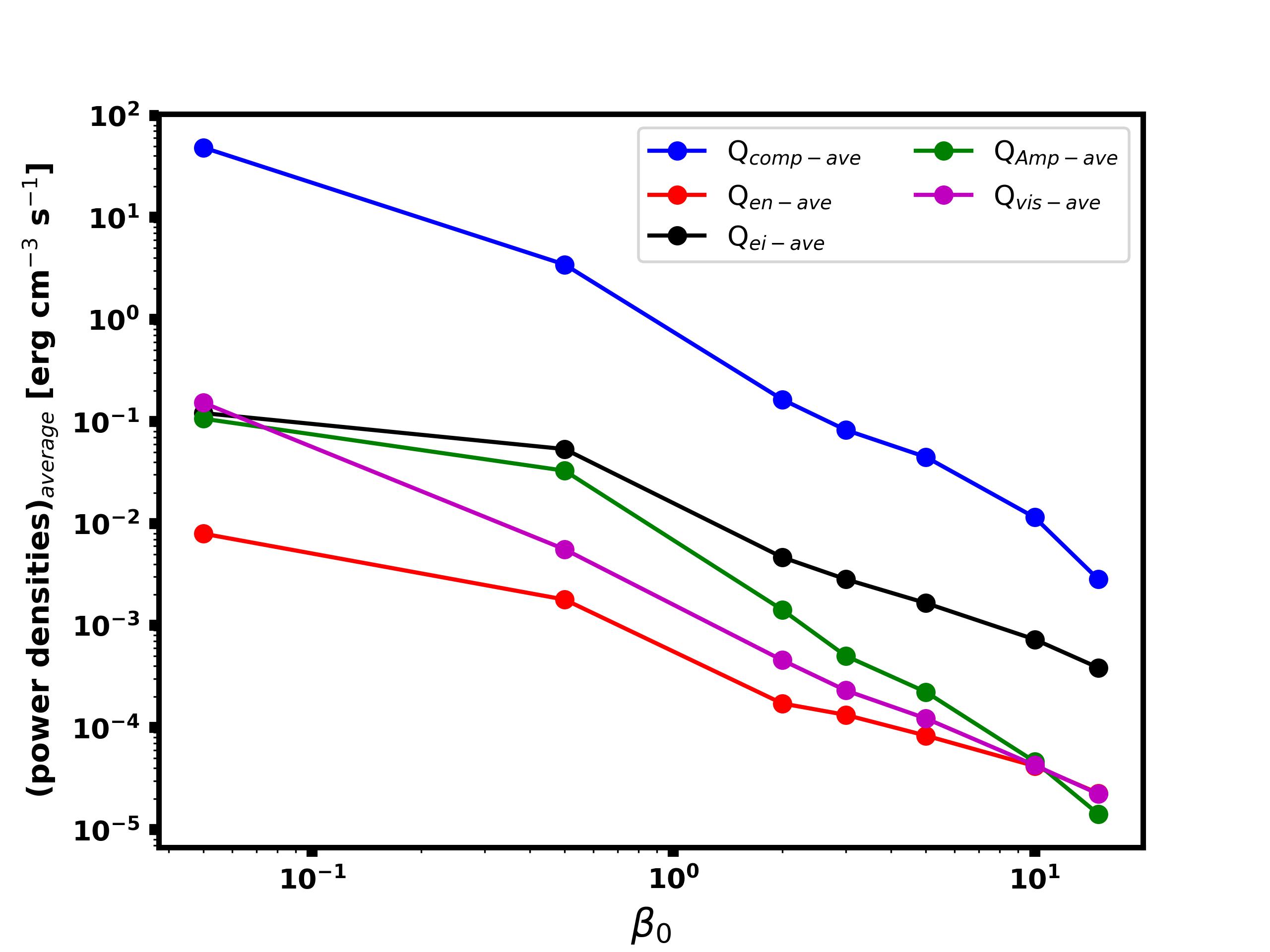}
\put(-160,175){\textbf{(d) Z = 1000 km}}
\end{minipage}
\begin{minipage}{0.494\textwidth}
\includegraphics[width=1.0\textwidth]{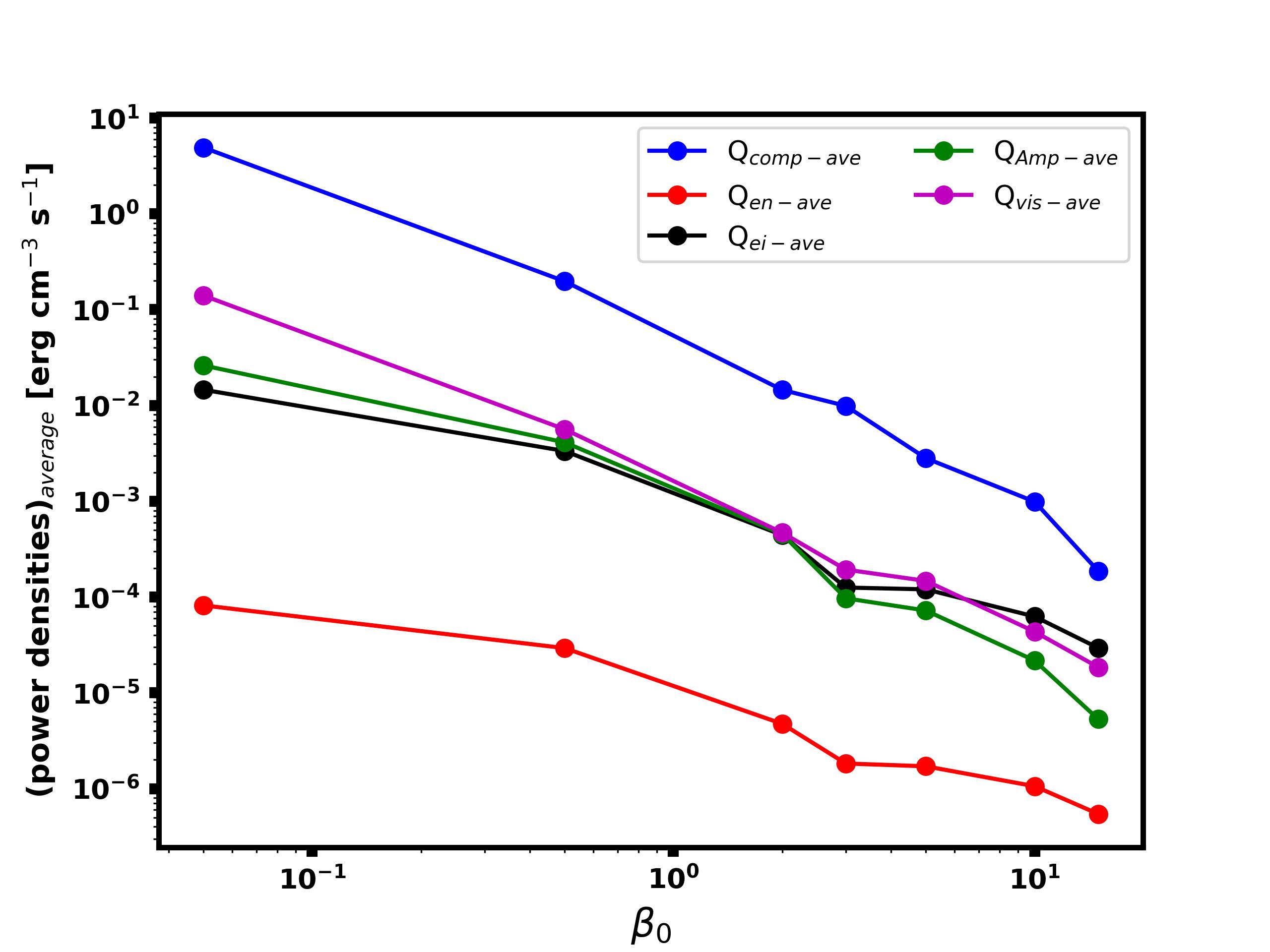}
\put(-160,175){\textbf{(e) Z = 1400 km}}
\end{minipage}
\begin{minipage}{0.494\textwidth}
\includegraphics[width=1.0\textwidth]{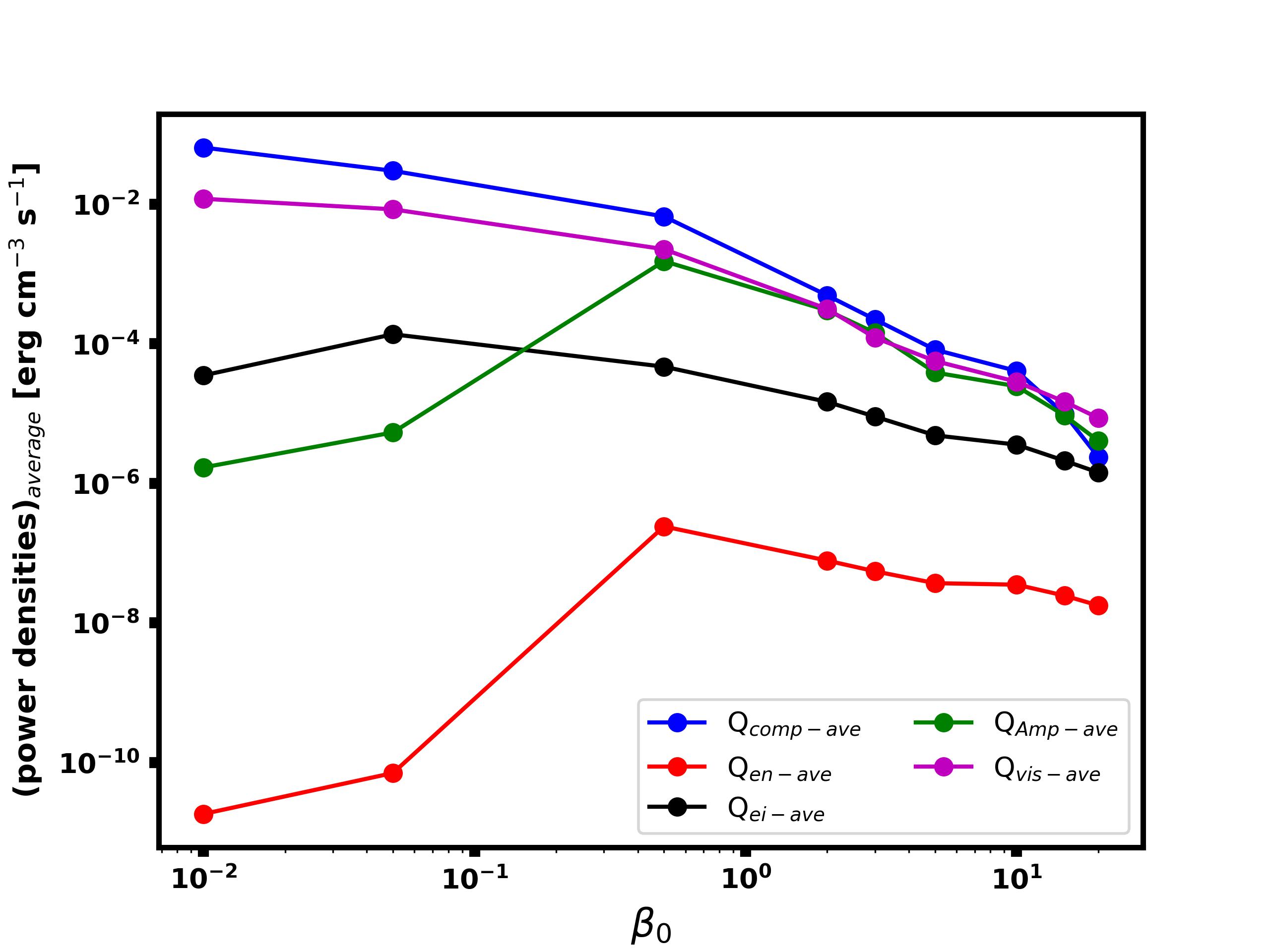}
\put(-160,175){\textbf{(f) Z = 2000 km}}
\end{minipage}
\caption{Evolutions of the average power densities contributed by various heating terms versus initial plasma-$\beta$ at Z = 400 km above the solar surface (panel a), Z = 600 km (panel b), Z = 800 km (panel c), Z = 1000 km (panel d), Z = 1400 km (panel e), and Z = 2000 km (panel f). The filled circles represent the average values calculated at different $\beta_0$ and the solid lines connect these data points.}
\label{fig_2}
\end{figure}

\section{Numerical Results}
\label{sec_III}

Figure~\ref{fig_1} shows the two-dimensional distribution of temperature (upper half panel) and divergence of velocity (lower half panel) with magnetic field lines (solid black lines) in different stages of the magnetic reconnection process at Z = 400 km (upper photosphere), with a lower initial plasma-$\beta$ ($\beta_{0}$ = 0.5) and a higher initial plasma-$\beta$ ($\beta_{0}$ = 10.0), respectively.
As time progresses, the aspect ratio of the Sweet-Parker type current sheet becomes too large, leading to tearing mode instability and fragmentation of the current sheet into multiple plasmoids (magnetic islands).
The coalescence of two or more magnetic islands causes the appearance of bigger islands.
Comparing the 2D distributions of the lower $\beta_{0}$ case (Figure~\ref{fig_1}(a)) and the higher $\beta_{0}$ case (Figure~\ref{fig_1}(b)), we can see a noticeable difference in the temperature and the $\nabla \cdot$ V distributions, as well as the number of plasmoids generated in the reconnection process.
The number of plasmoids varies with $\beta_{0}$, simulation with a lower $\beta_{0}$ (stronger magnetic field) generates more plasmoids.
Plasma is heated to significantly higher temperature in a lower $\beta_{0}$ case when compared to their corresponding higher $\beta_{0}$ case.
The lower half panels tell us that the divergence of the plasma velocity is more pronounced in the simulation with $\beta_{0}$ = 0.5, plasma compression is prominent ($\nabla \cdot V < 0$).
Many compression fronts are also observed in the inflow region, which are slow shocks generated by plasmoids. Such slow-mode shocks are playing a role in boosting the magnetic reconnection rate~\citep{li2012slow,wu2017shock,zafar2024unraveling}.
While for $\beta_{0}$ = 10.0 (Figure~\ref{fig_1}(b)), we cannot see obvious signatures of plasma compression ($\nabla \cdot V < 0$) or expansion ($\nabla \cdot V > 0$) inside the current sheet or in the inflow region.
Hence, these findings suggest that plasma compression is an important characteristic in lower $\beta_{0}$ scenarios.

\begin{figure}
\centering
\begin{minipage}{0.494\textwidth}
\includegraphics[width=1.0\textwidth]{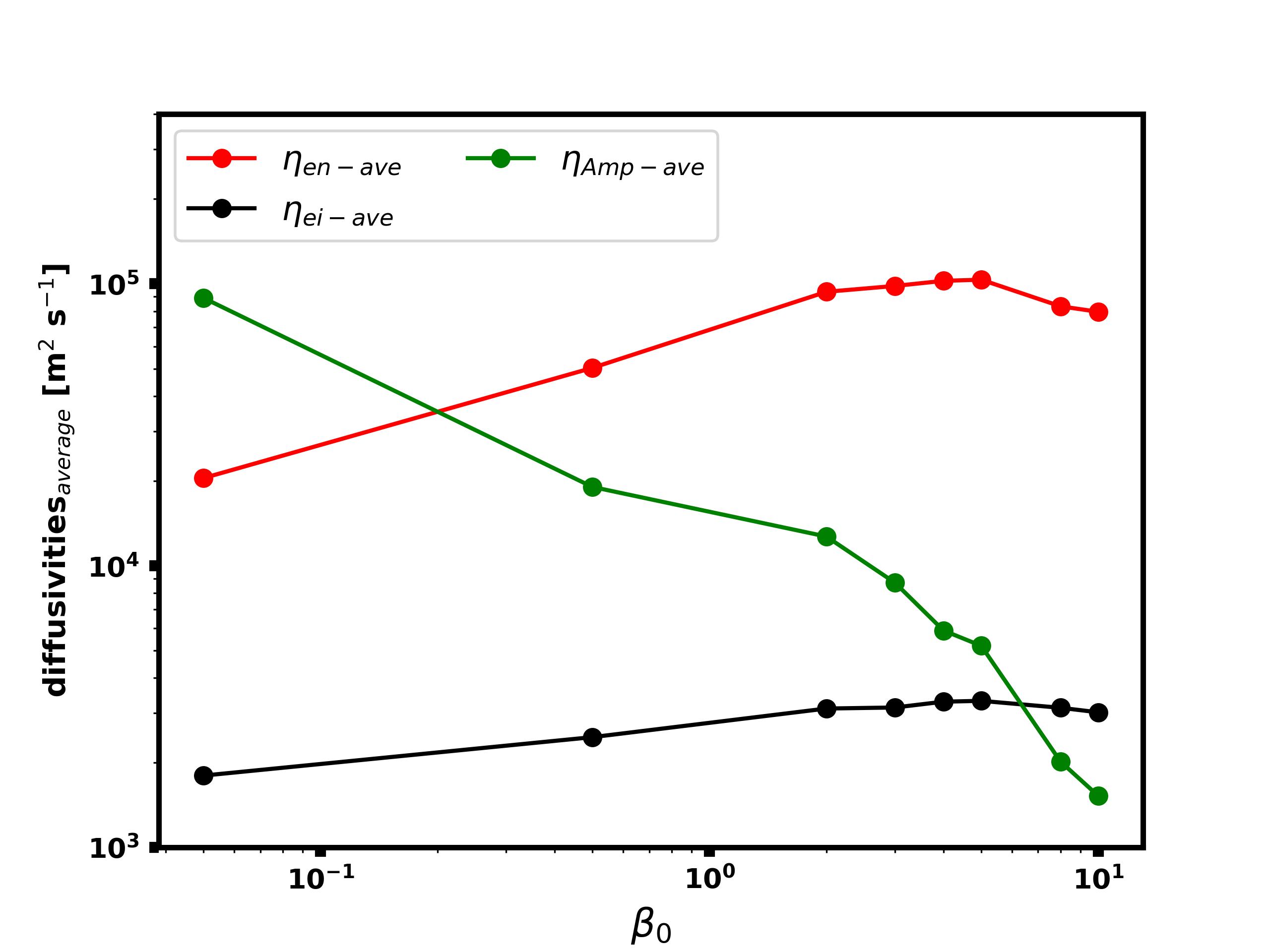}
\put(-160,175){\textbf{(a) Z = 400 km}}
\end{minipage}
\begin{minipage}{0.494\textwidth}
\includegraphics[width=1.0\textwidth]{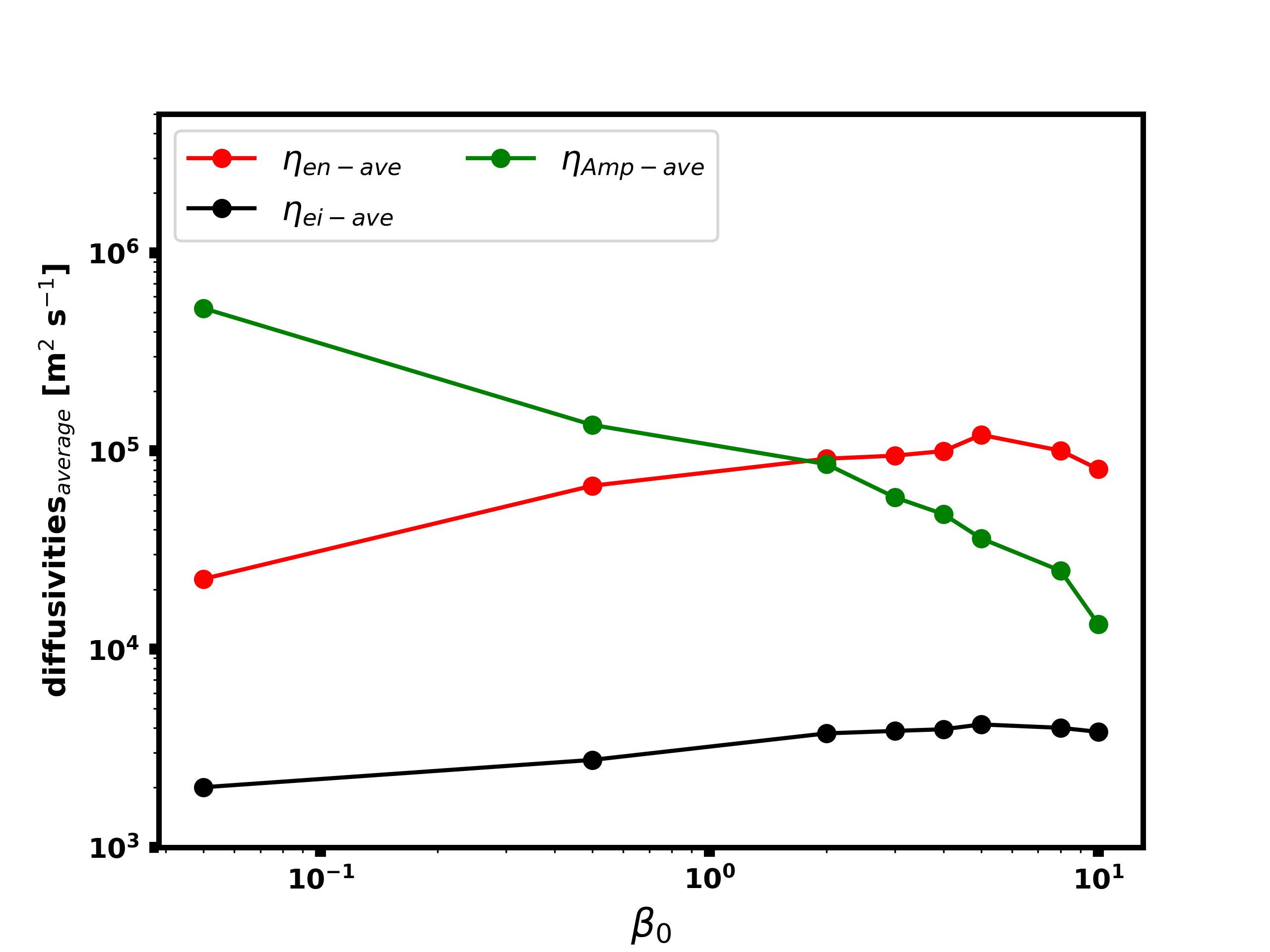}
\put(-160,175){\textbf{(b) Z = 600 km}}
\end{minipage}
\begin{minipage}{0.494\textwidth}
\includegraphics[width=1.0\textwidth]{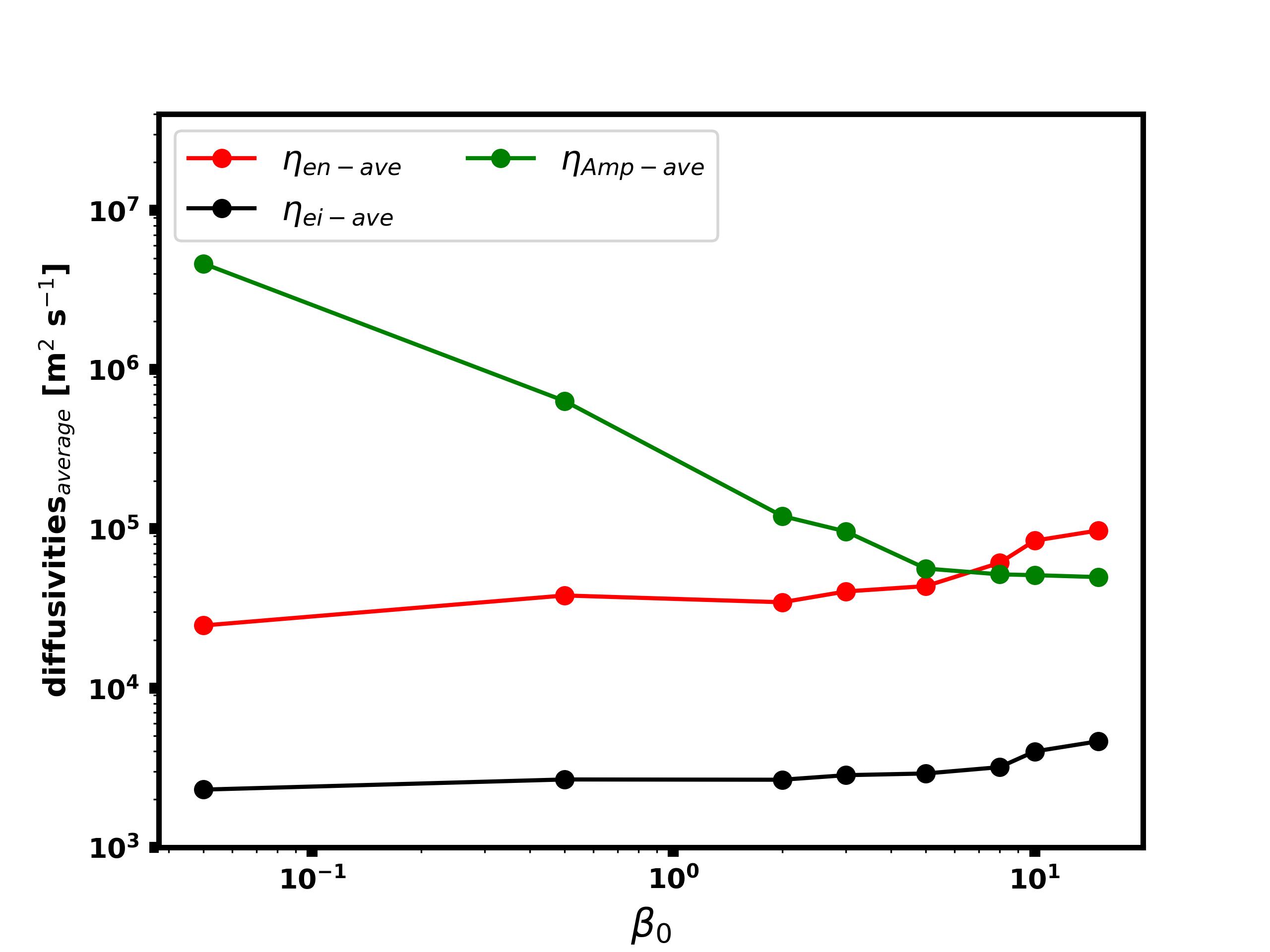}
\put(-160,175){\textbf{(c) Z = 800 km}}
\end{minipage}
\begin{minipage}{0.494\textwidth}
\includegraphics[width=1.0\textwidth]{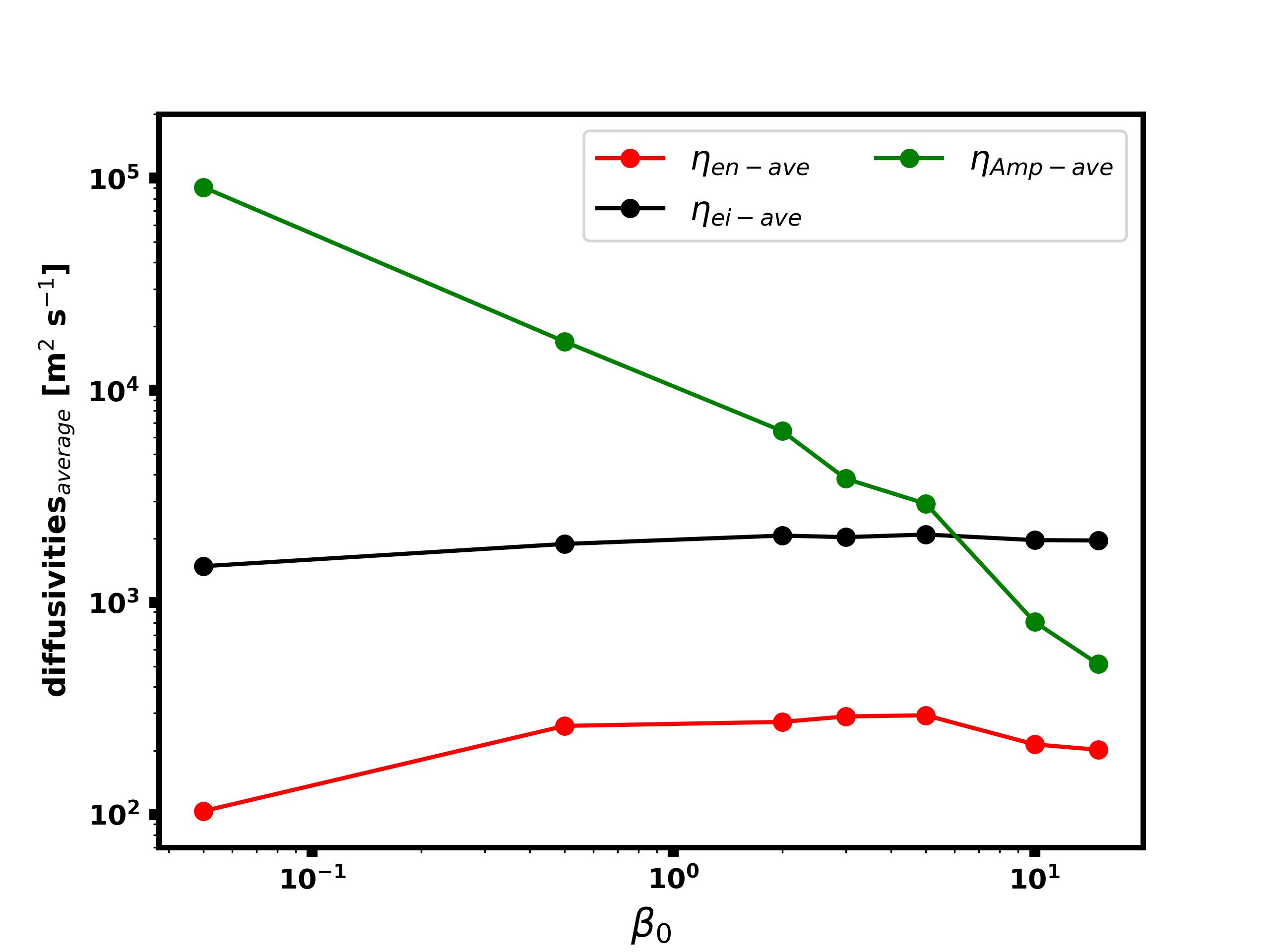}
\put(-160,175){\textbf{(d) Z = 1000 km}}
\end{minipage}
\begin{minipage}{0.494\textwidth}
\includegraphics[width=1.0\textwidth]{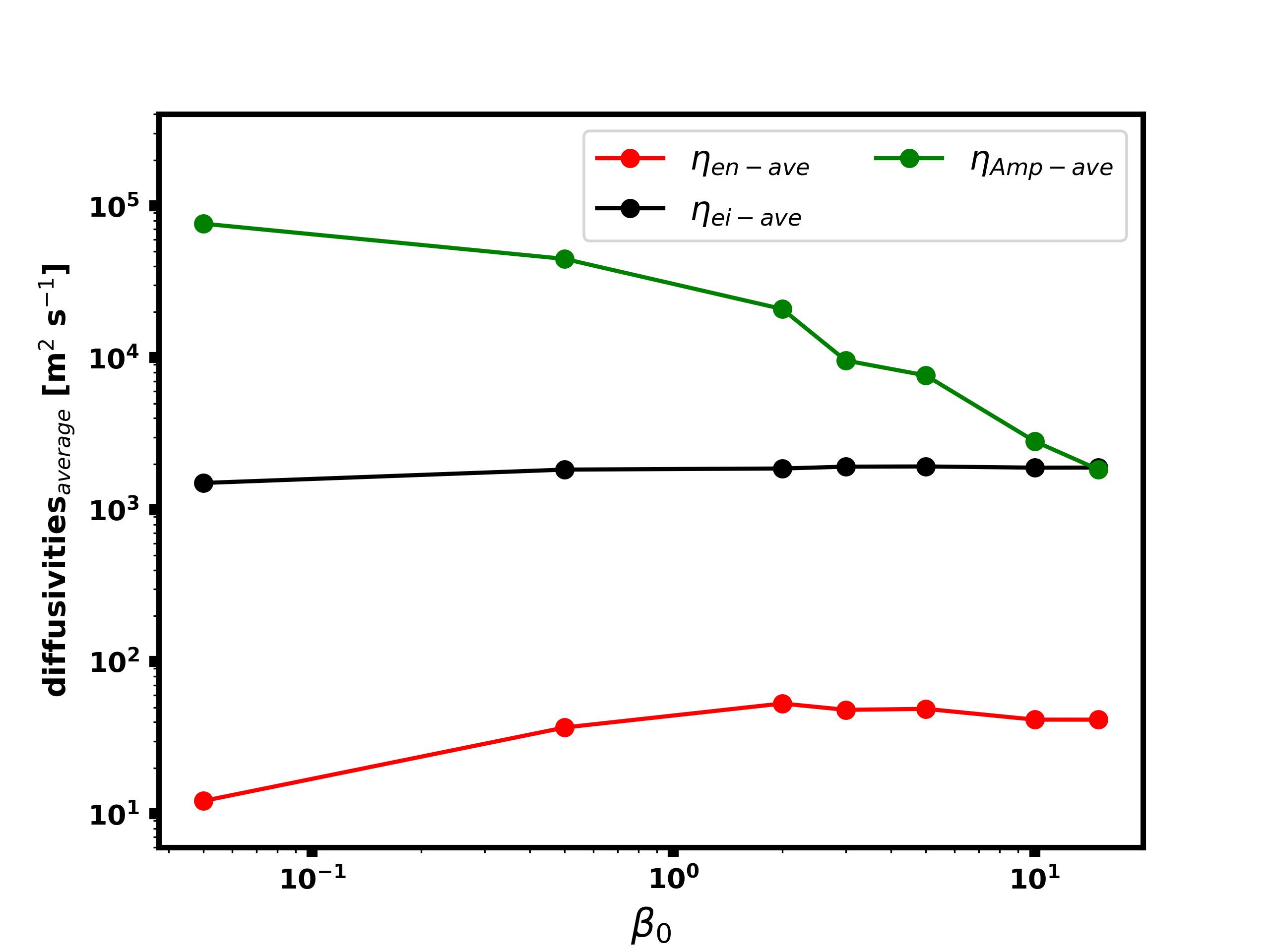}
\put(-160,175){\textbf{(e) Z = 1400 km}}
\end{minipage}
\begin{minipage}{0.494\textwidth}
\includegraphics[width=1.0\textwidth]{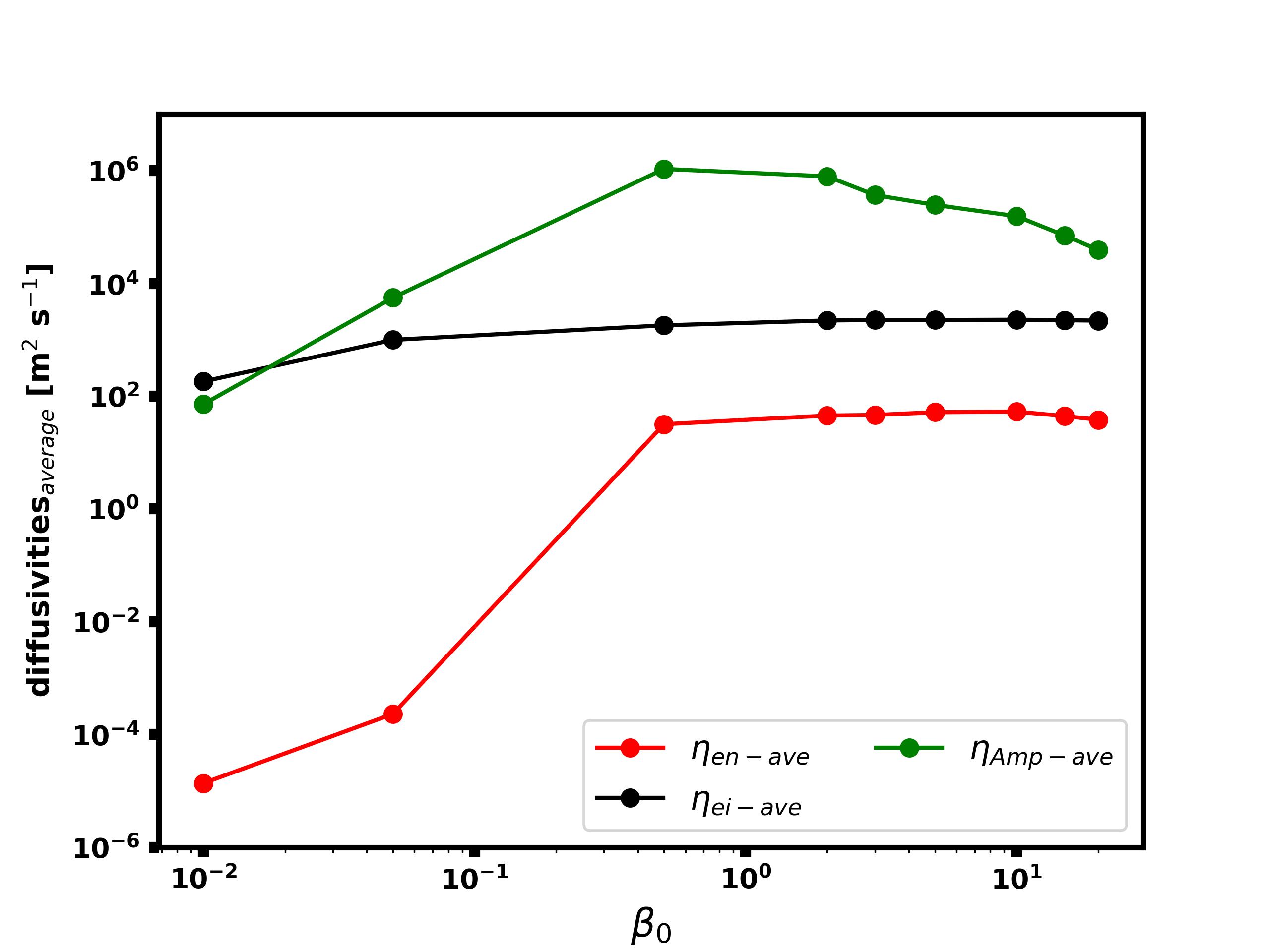}
\put(-160,175){\textbf{(f) Z = 2000 km}}
\end{minipage}
\caption{The evolution of the average diffusion coefficients ($\eta_{en-ave}$, $\eta_{ei-ave}$, $\eta_{Amp-ave}$) with $\beta_{0}$ at different altitudes in the lower solar atmosphere (a) Z = 400 km, (b) Z = 600 km, (c) Z = 800 km, (d) Z = 1000 km, (e) Z = 1400 km, and (f) Z = 2000 km. The filled circles represent the average values calculated at different $\beta_0$ and the solid lines connect these data points.}
\label{fig_3}
\end{figure}

\begin{figure}
\centering
\begin{minipage}{0.494\textwidth}
\includegraphics[width=1.0\textwidth]{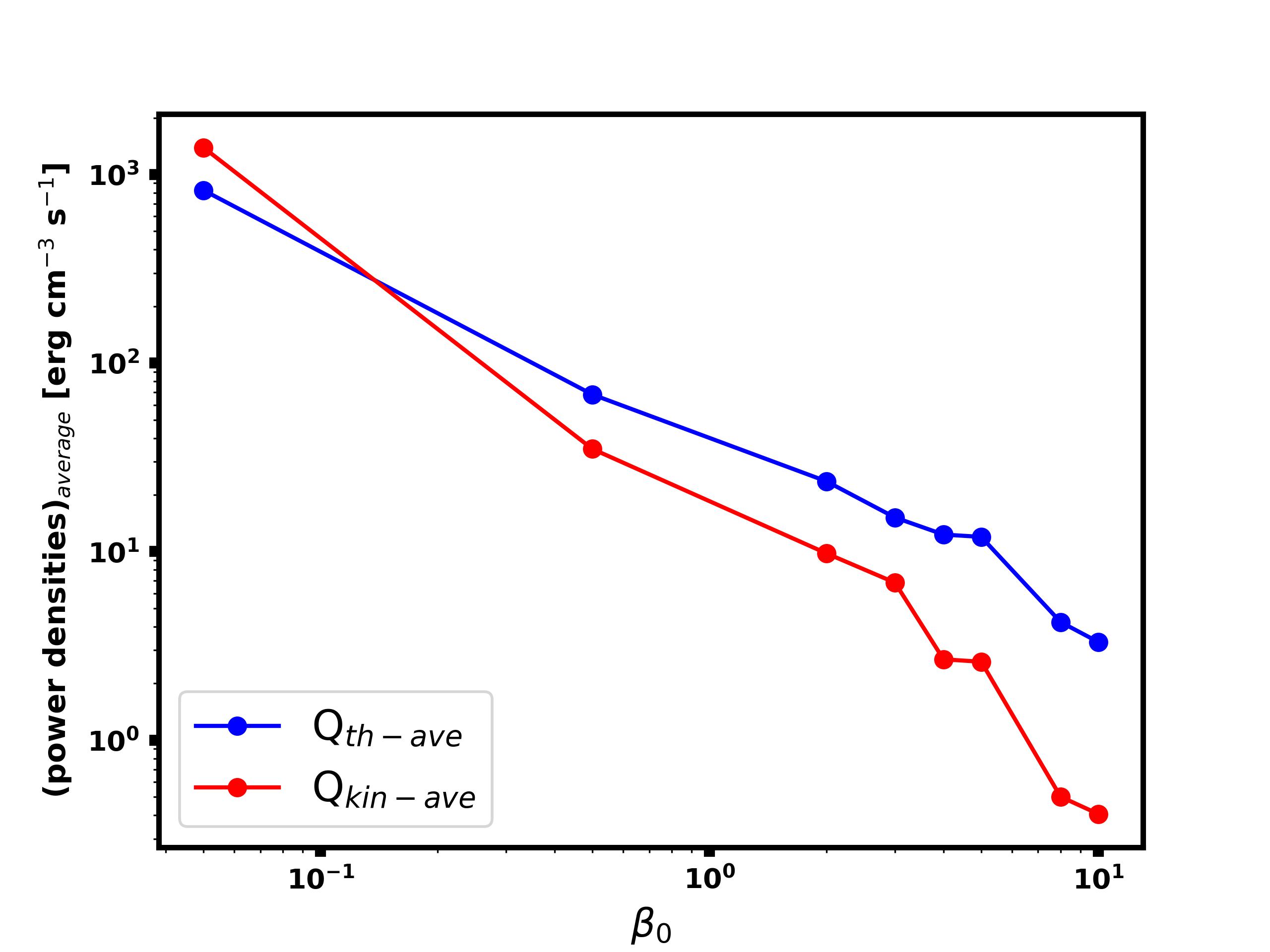}
\put(-160,175){\textbf{(a) Z = 400 km}}
\end{minipage}
\begin{minipage}{0.494\textwidth}
\includegraphics[width=1.0\textwidth]{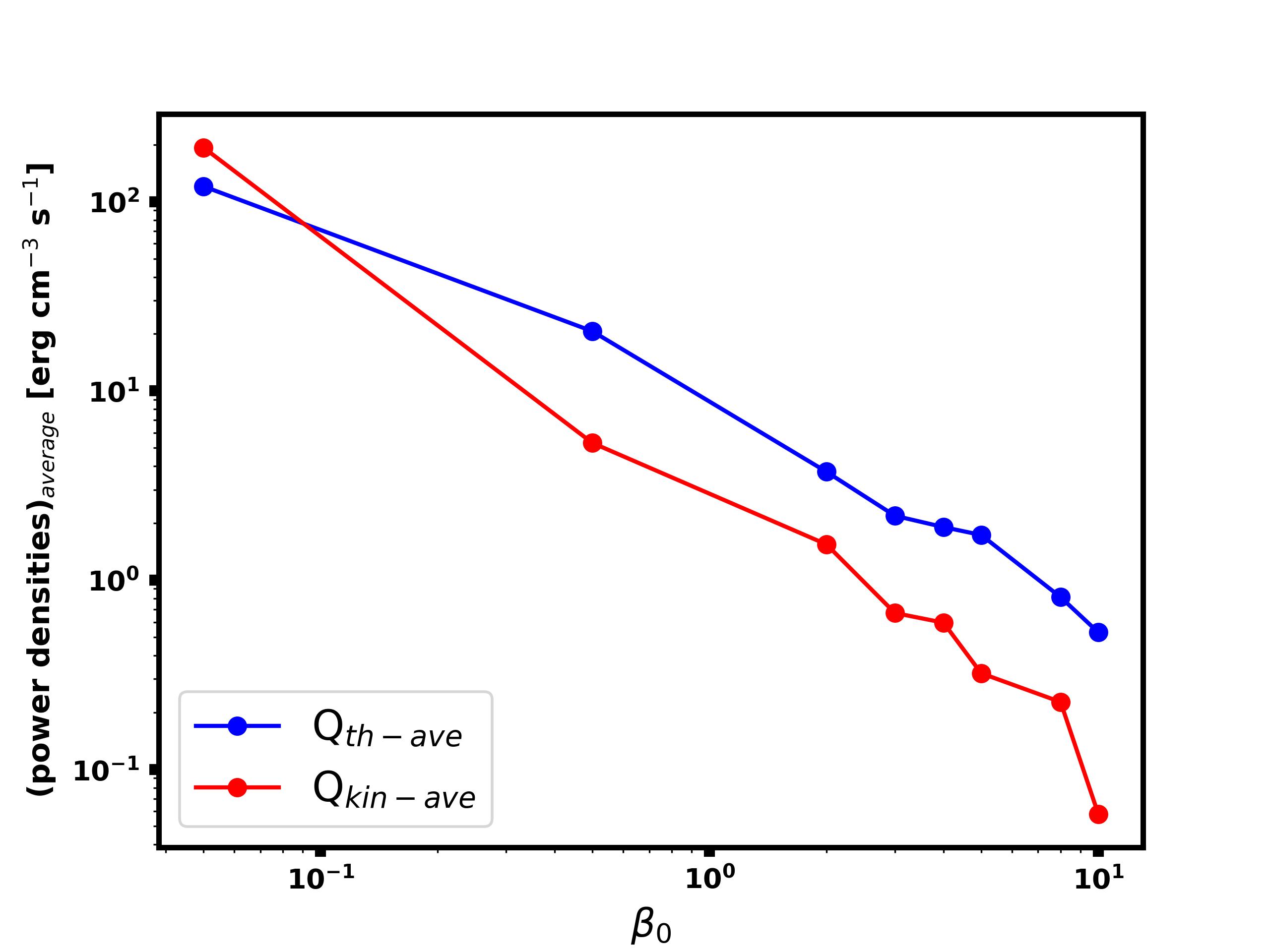}
\put(-160,175){\textbf{(b) Z = 600 km}}
\end{minipage}
\begin{minipage}{0.494\textwidth}
\includegraphics[width=1.0\textwidth]{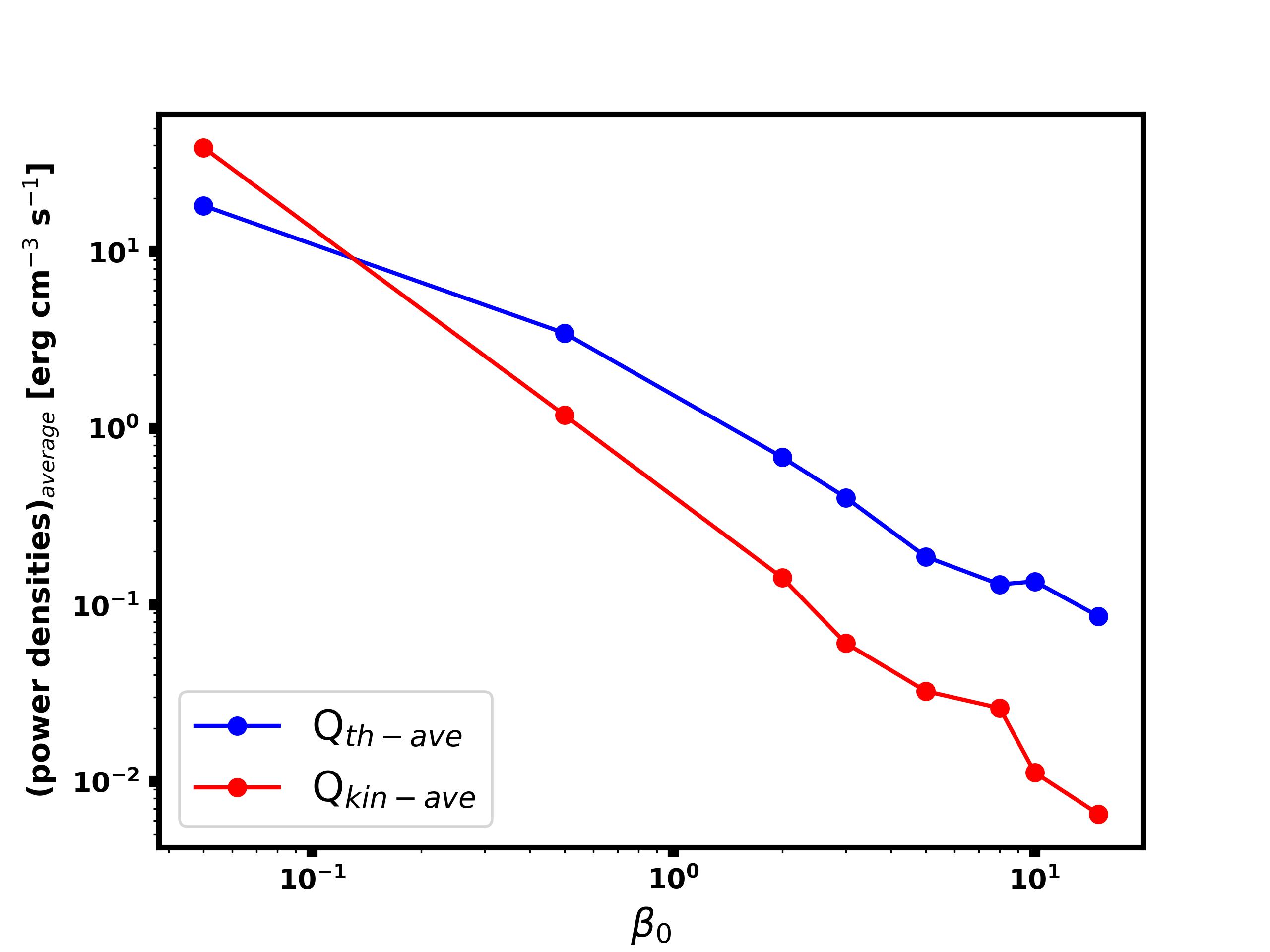}
\put(-160,175){\textbf{(c) Z = 800 km}}
\end{minipage}
\begin{minipage}{0.494\textwidth}
\includegraphics[width=1.0\textwidth]{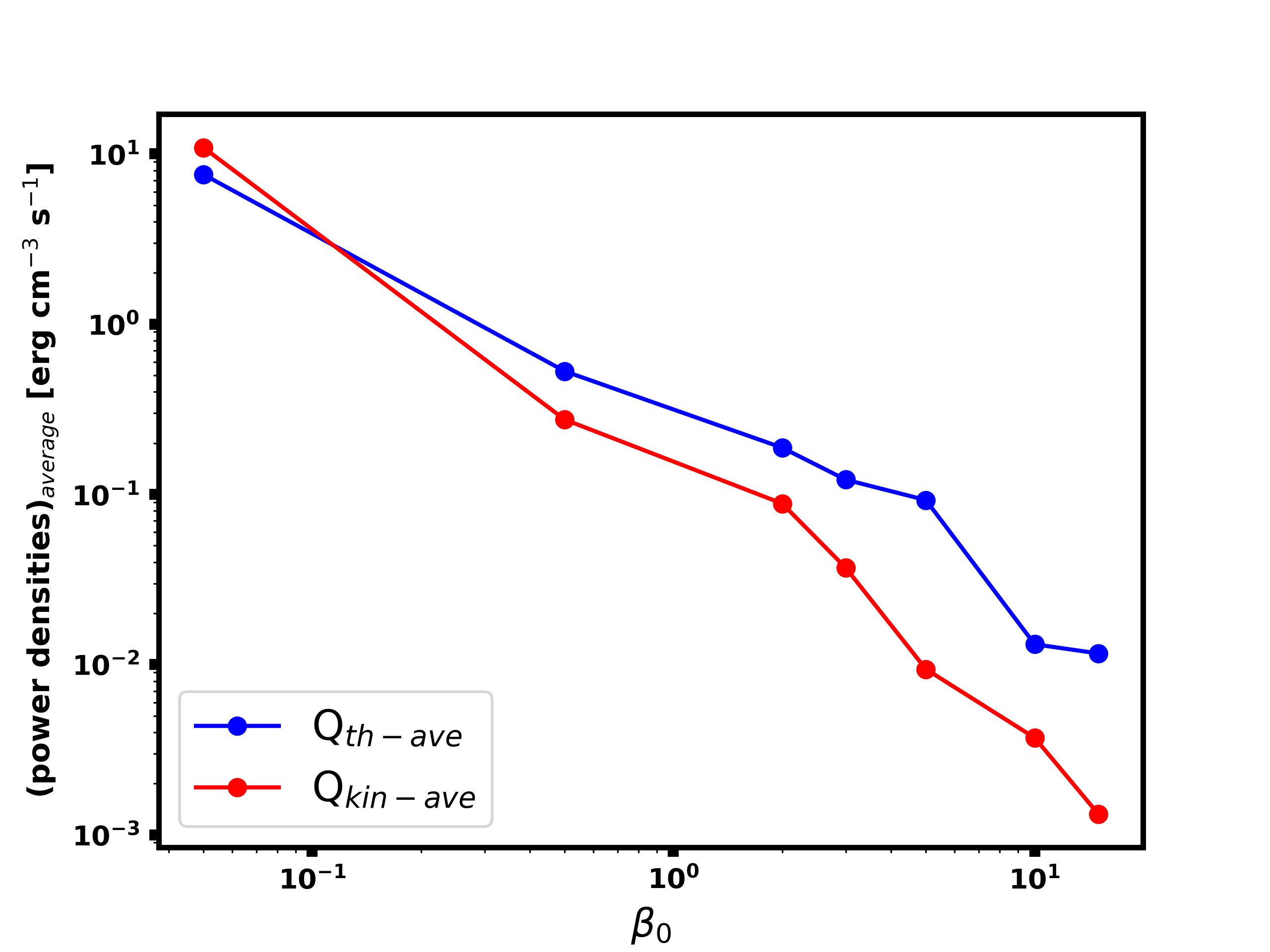}
\put(-160,175){\textbf{(d) Z = 1000 km}}
\end{minipage}
\begin{minipage}{0.494\textwidth}
\includegraphics[width=1.0\textwidth]{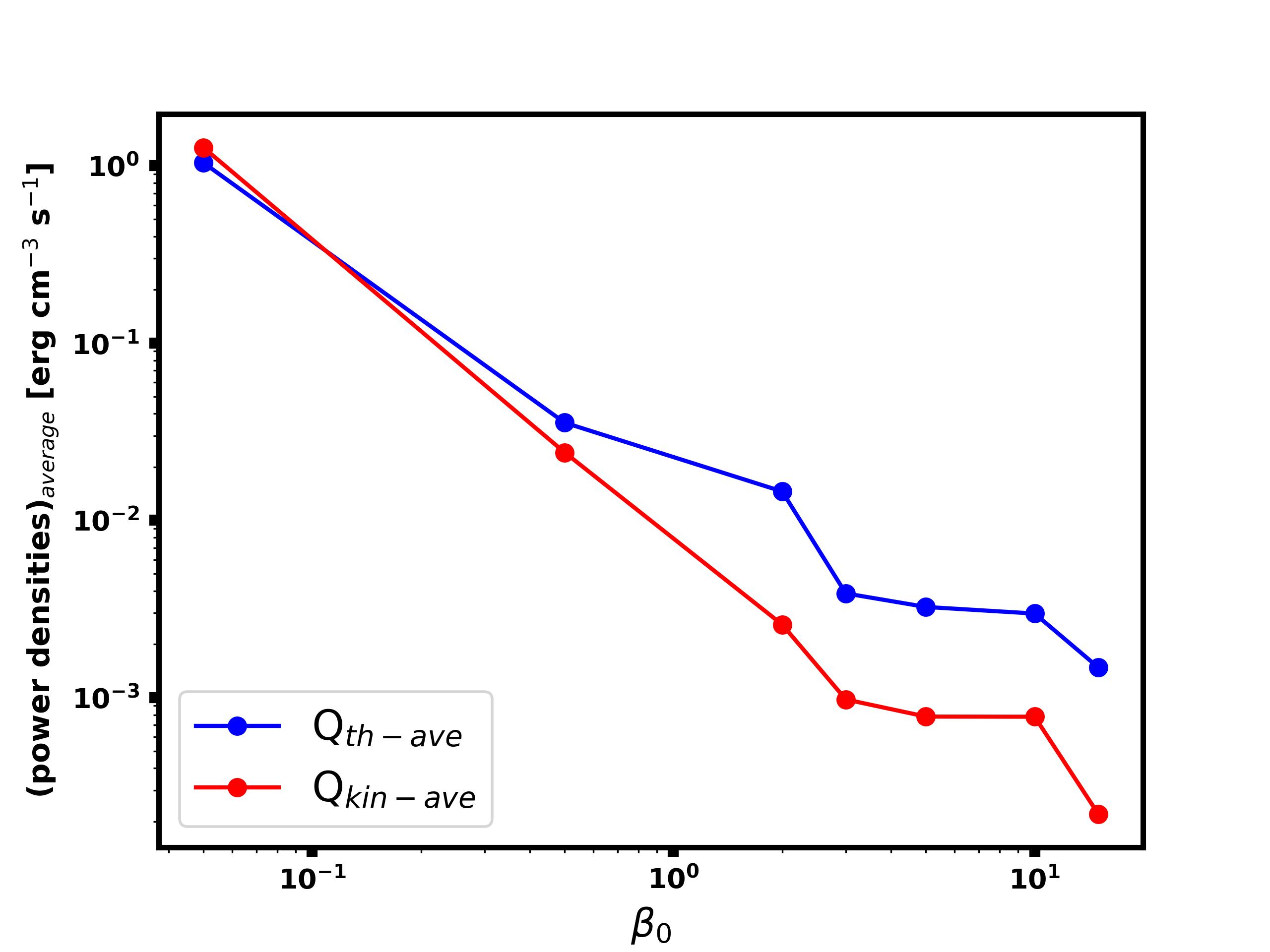}
\put(-160,175){\textbf{(e) Z = 1400 km}}
\end{minipage}
\begin{minipage}{0.494\textwidth}
\includegraphics[width=1.0\textwidth]{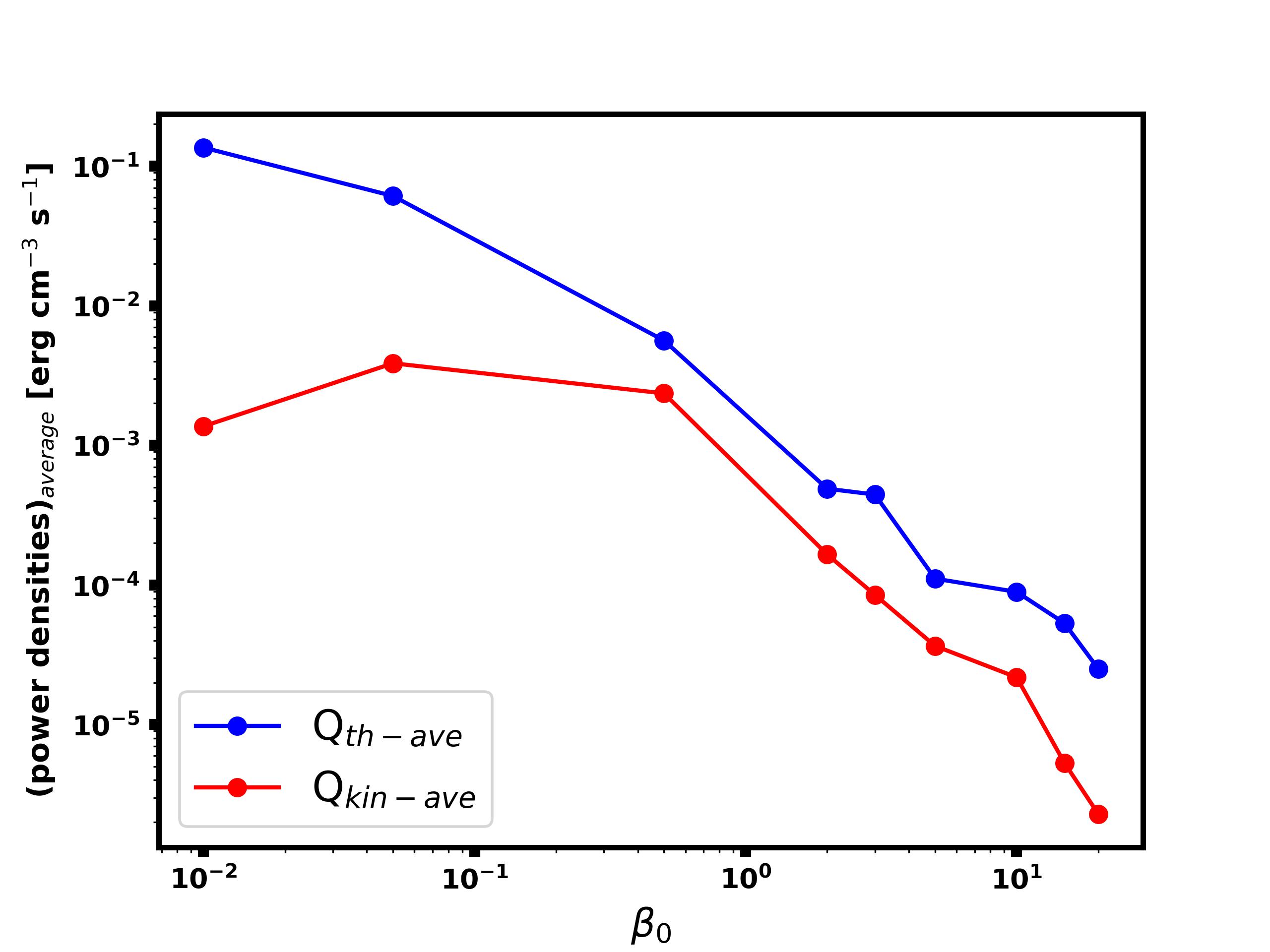}
\put(-160,175){\textbf{(f) Z = 2000 km}}
\end{minipage}
\caption{Evolutions of the average thermal power density (Q$_{th-ave}$) and kinetic power density (Q$_{kin-ave}$) versus initial plasma-$\beta$ at Z = 400 km above the solar surface (panel a), Z = 600 km (panel b), Z = 800 km (panel c), Z = 1000 km (panel d), Z = 1400 km (panel e), and Z = 2000 km (panel f). The filled circles represent the average values calculated at different $\beta_0$ and the solid lines connect these data points.}
\label{fig_5}
\end{figure}

In the MHD model, the thermal energy density equation is:
\begin{eqnarray}
\frac{d e_{th}}{dt} = -p \nabla \cdot \mathbf{v} + \frac{1}{2 \xi} Tr(\tau^{2}_{S}) + \frac{\eta_{ei,en}}{\mu_{0}} |\nabla \times \mathbf{B}|^{2} + \frac{\eta_{AD}}{\mu_{0}^{2}} |\mathbf{B} \times (\nabla \times \mathbf{B})|^{2} + Q_{rad}. 
\label{eq:en_density}
\end{eqnarray}
where $e_{th}$ denotes the thermal energy density.
The first four terms on the right side of the thermal density equation represent various heating terms, including heating caused by compressional effect, Q$_{comp}$ = $-p \nabla \cdot \mathbf{v}$, viscous heating, Q$_{vis}$ = $\frac{1}{2 \xi} Tr(\tau^{2}_{S})$, Joule heating contributed by electrons-ions and electrons-neutrals collisions, Q$_{ei,en}$ =  $\frac{\eta_{ei,en}}{\mu_{0}} |\nabla \times \mathbf{B}|^{2}$ and heating by ambipolar diffusion, Q$_{Amp}$ =  $\frac{\eta_{AD}}{\mu_{0}^{2}} |\mathbf{B} \times (\nabla \times \mathbf{B})|^{2}$. The last term, $Q_{rad}$, measures the loss of thermal energy due to radiative cooling.
These heating terms, kinetic energy (Q$_{kin}$), and thermal energy (Q$_{th}$) are computed using the same approach as reported in Ni et al.~\citep{ni2022plausibility}.
The averaged values contributed by different heating terms and kinetic and thermal energies for all the numerical experiments presented in this study are determined throughout the reconnection region within the simulation box for $0 \leq x \leq L_{0}$ and $-0.05L_{0} \leq y \leq 0.05L_{0}$.
The data points for the heating scenarios, diffusion coefficients, and thermal and kinetic power densities at different $\beta_{0}$ in Figures~\ref{fig_2}, \ref{fig_3} and \ref{fig_5} respectively, are evaluated by taking the average values after the simulations reach the saturation stage.

\begin{figure}
\centering
\begin{minipage}{0.494\textwidth}
\includegraphics[width=1.0\textwidth]{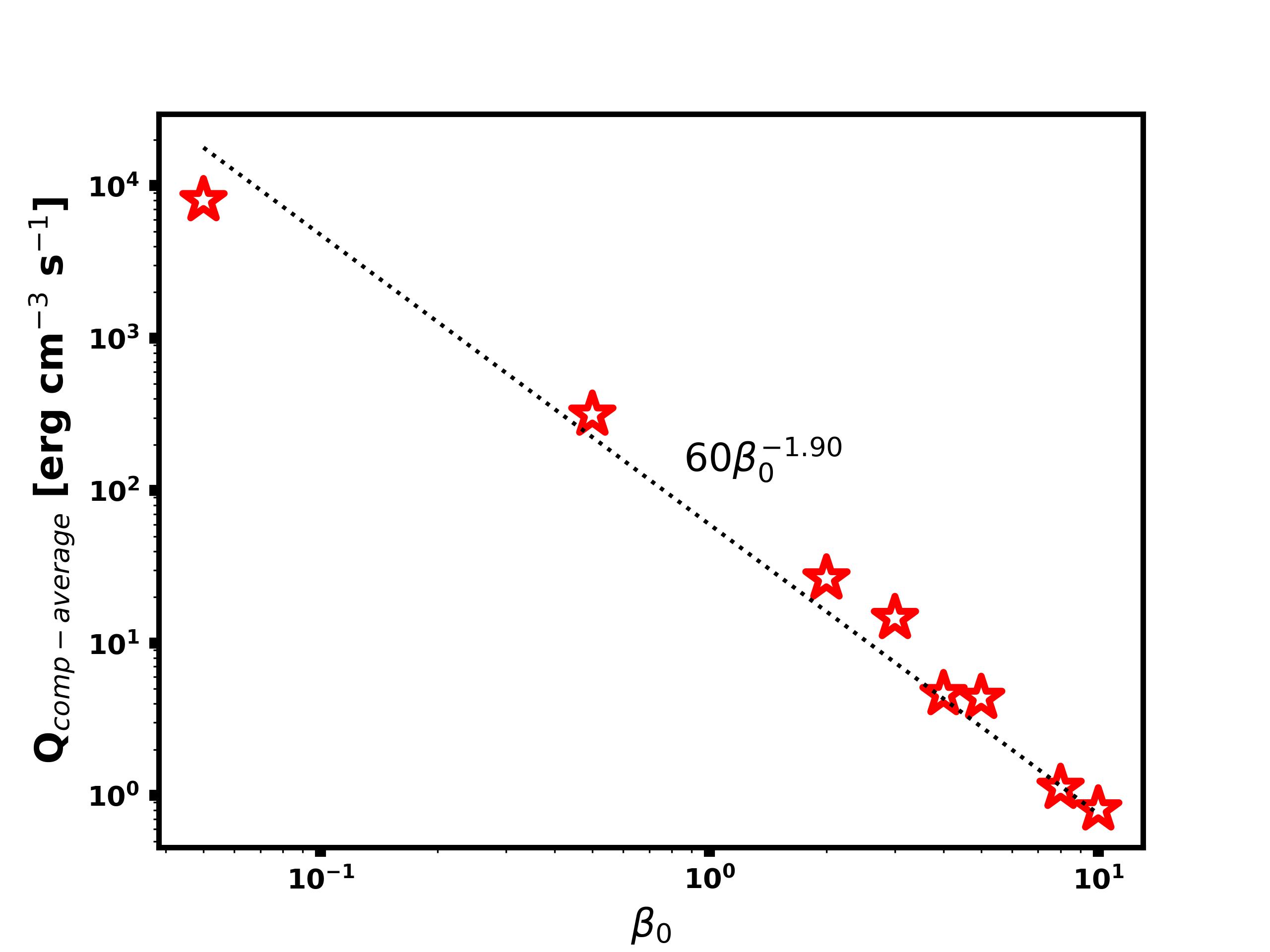}
\put(-160,175){\textbf{(a) Z = 400 km}}
\end{minipage}
\begin{minipage}{0.494\textwidth}
\includegraphics[width=1.0\textwidth]{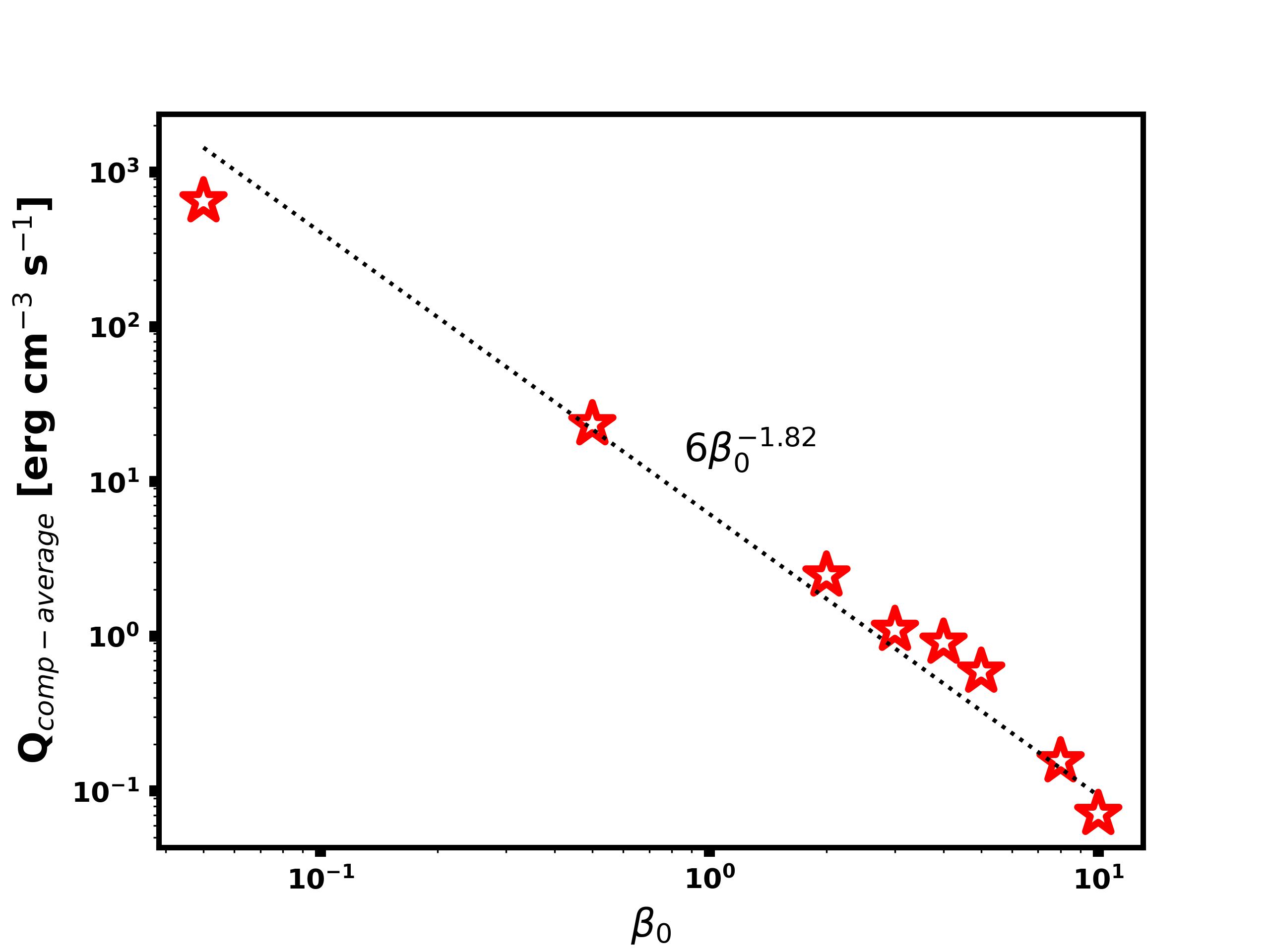}
\put(-160,175){\textbf{(b) Z = 600 km}}
\end{minipage}
\begin{minipage}{0.494\textwidth}
\includegraphics[width=1.0\textwidth]{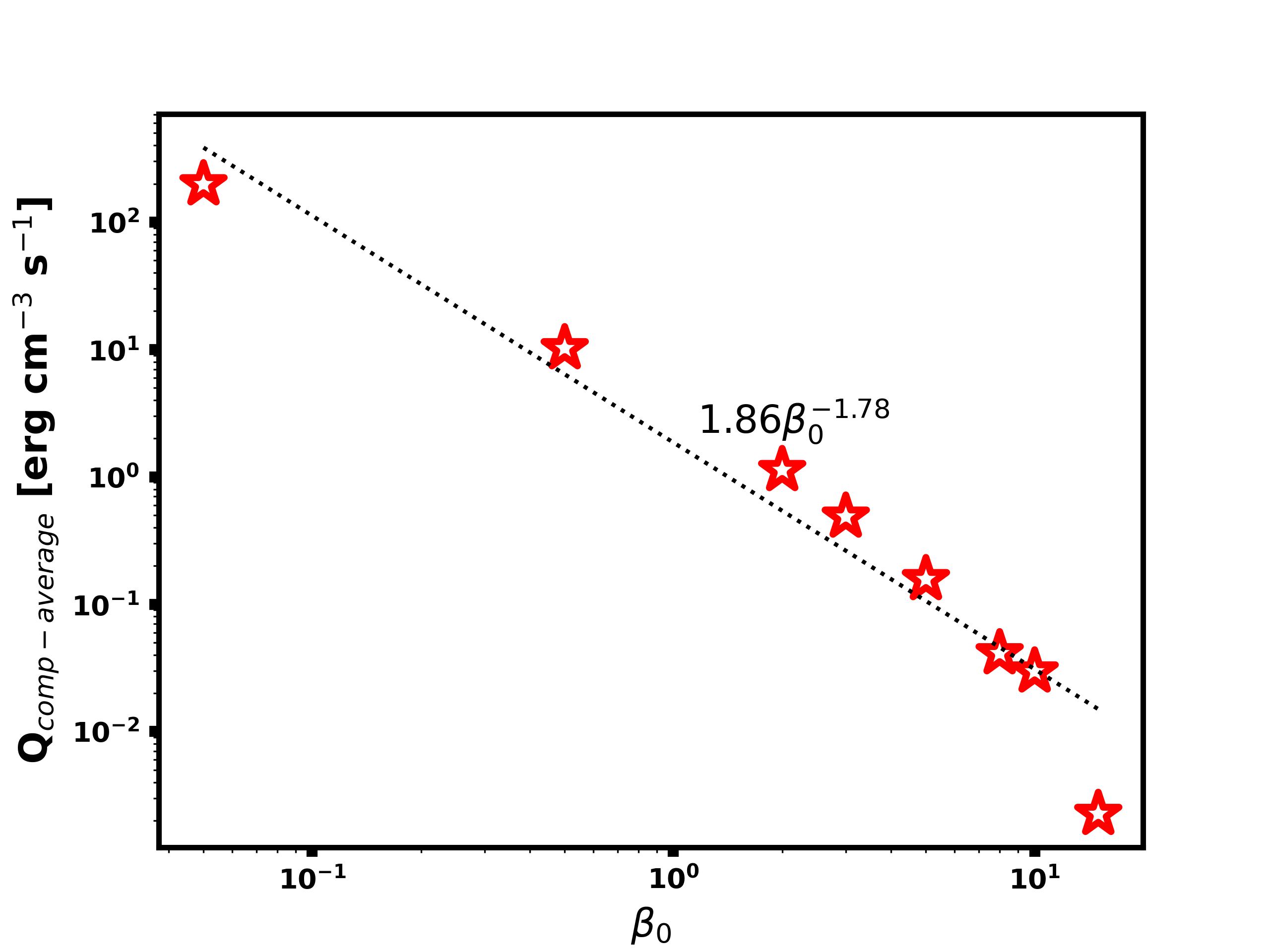}
\put(-160,175){\textbf{(c) Z = 800 km}}
\end{minipage}
\begin{minipage}{0.494\textwidth}
\includegraphics[width=1.0\textwidth]{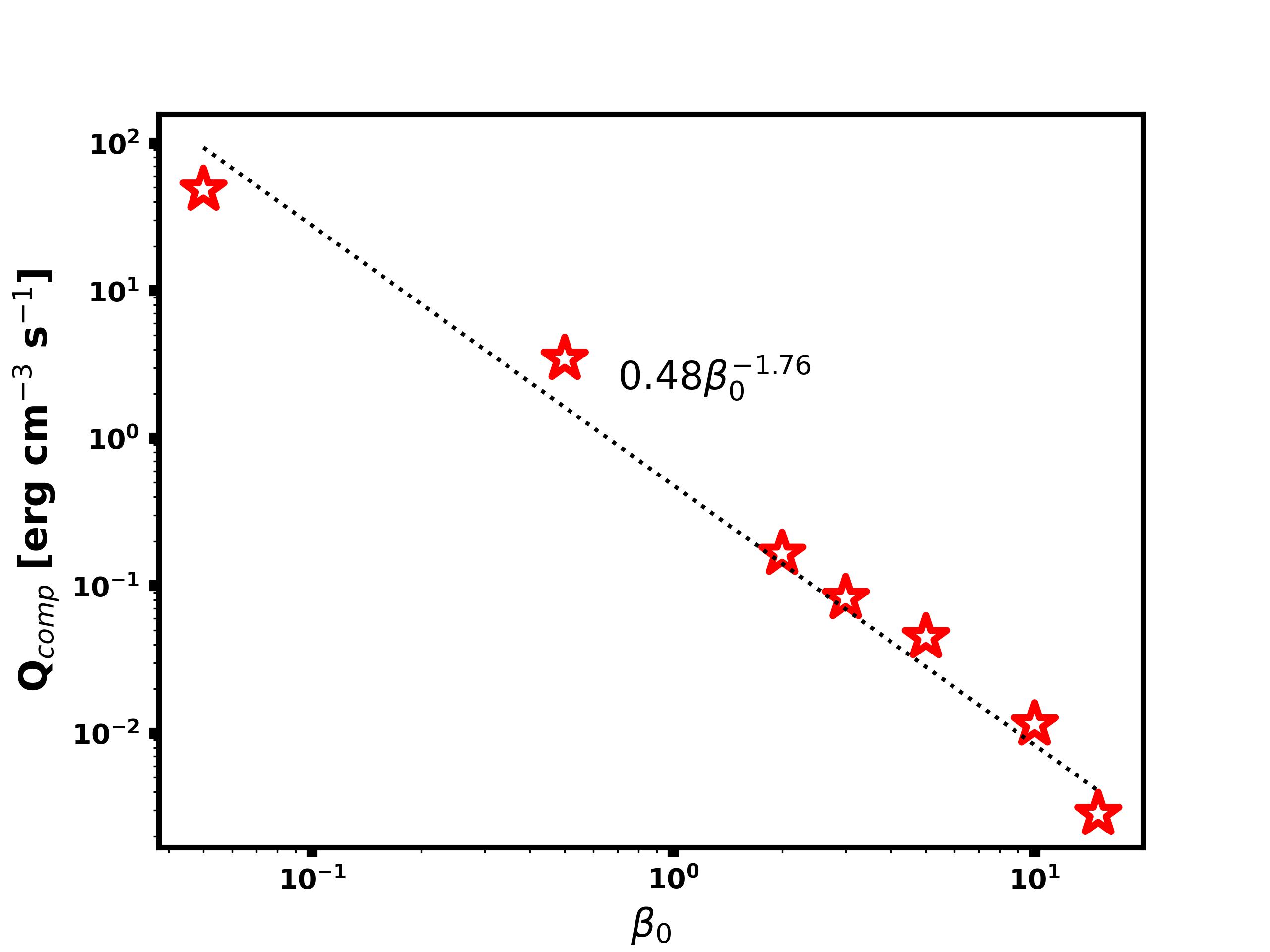}
\put(-160,175){\textbf{(d) Z = 1000 km}}
\end{minipage}
\begin{minipage}{0.494\textwidth}
\includegraphics[width=1.0\textwidth]{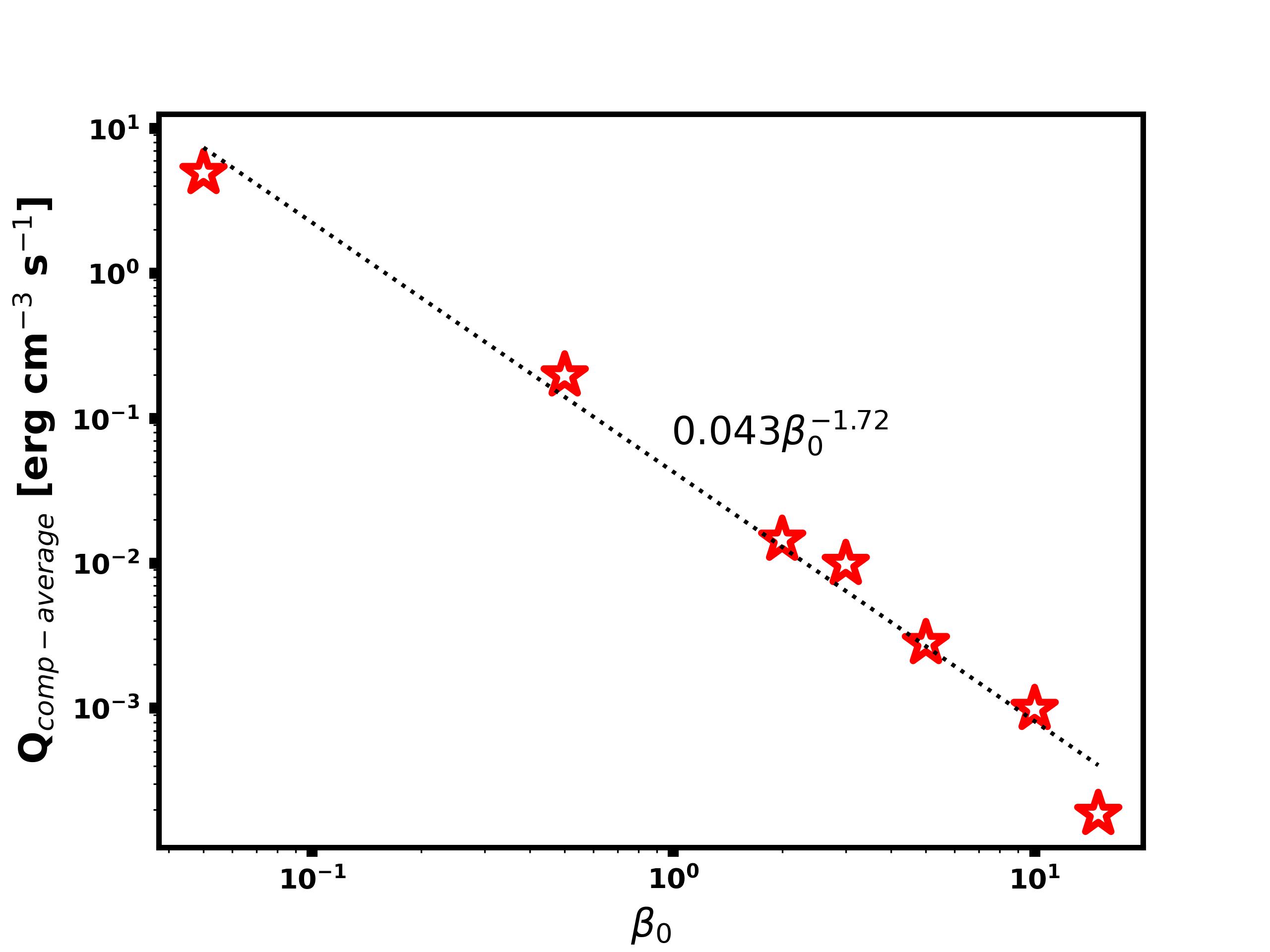}
\put(-160,175){\textbf{(e) Z = 1400 km}}
\end{minipage}
\begin{minipage}{0.494\textwidth}
\includegraphics[width=1.0\textwidth]{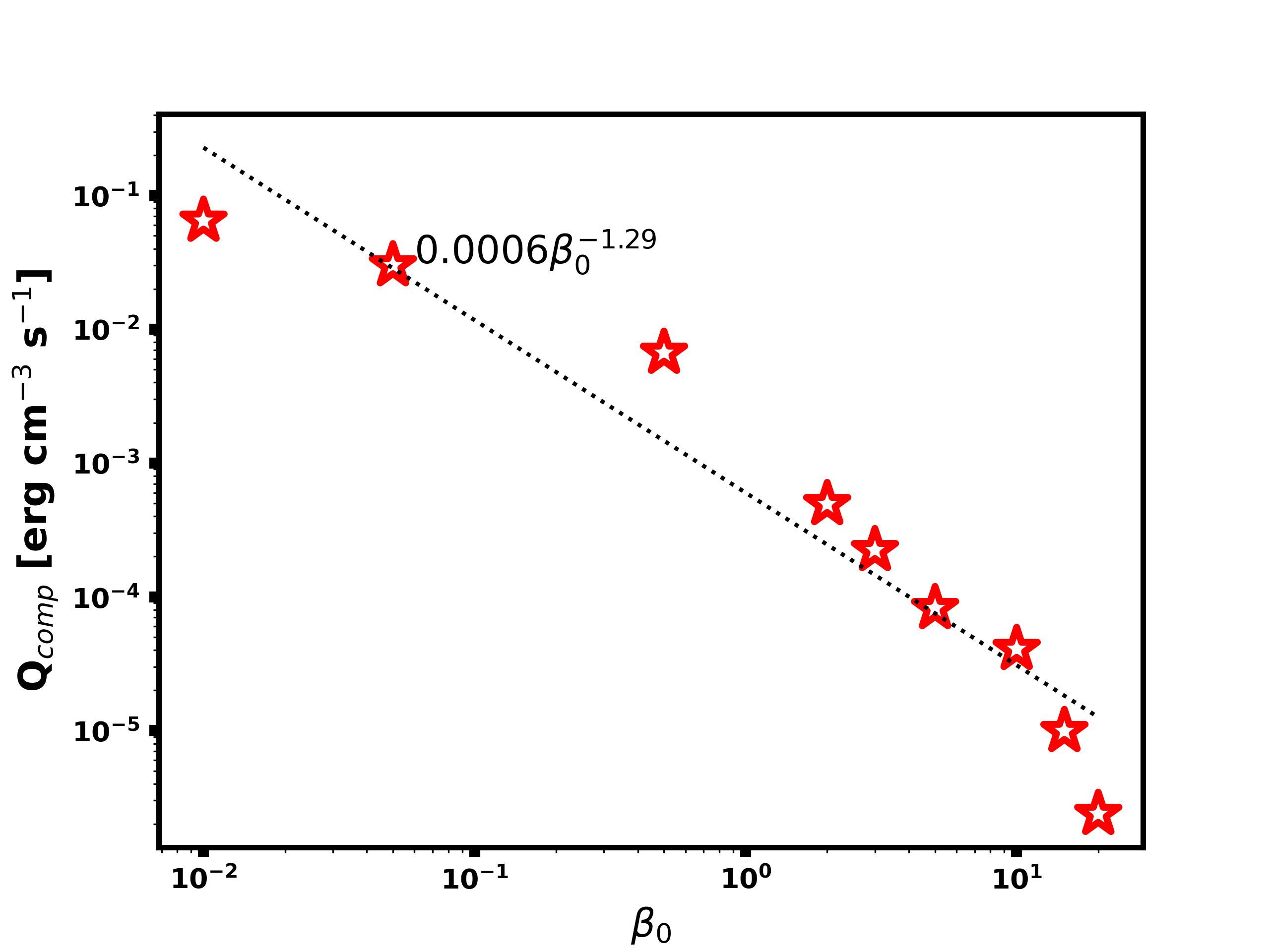}
\put(-160,175){\textbf{(f) Z = 2000 km}}
\end{minipage}
\caption{The evolution of the average compressional heating (Q$_{comp-average}$) with $\beta_{0}$ at different altitudes in the lower solar atmosphere (a) Z = 400 km, (b) Z = 600 km, (c) Z = 800 km, (d) Z = 1000 km, (e) Z = 1400 km, and (f) Z = 2000 km. The back dotted lines demonstrate the power law fits.}
\label{fig_4}
\end{figure}

Figures~\ref{fig_2}(a)-\ref{fig_2}(f) demonstrate that Q$_{comp-average}$ is larger than the average values of the other heating terms in lower $\beta_{0}$ cases, indicating that compression heating (Q$_{comp}$) is the primary candidate to heat the plasma inside the reconnection region, particularly in strong magnetic field events at all different altitudes in the lower solar atmosphere.
Q$_{comp-average}$ decreases with $\beta_{0}$ in all simulation cases.
Whereas, the joule heating resulting from electron and neutral collisions (Q$_{en}$) responds differently with the modification of the initial plasma-$\beta$ below the middle chromosphere (see Figures~\ref{fig_2}(a)-\ref{fig_2}(c)).
Initially, Q$_{en-average}$ decreases with $\beta_{0}$, but later increases and finally decreases again with $\beta_0$.
Q$_{en-average}$ at Z = 400 km and 600 km begins to increase when $\beta_{0} >$ 0.5, and the difference in amplitude between Q$_{comp-average}$ and Q$_{en-average}$ decreases with increasing $\beta_{0}$.
Finally, when $\beta_{0}$ is around 8.0, the value of Q$_{en-average}$ exceeds Q$_{comp-average}$, indicating that compression heating is no longer the dominant plasma heating mechanism in such higher $\beta$ reconnection cases (Figures~\ref{fig_2}(a) and \ref{fig_2}(b)).
A similar trend is observed at Z = 800 km case (Figure~\ref{fig_2}(c)).
These results emphasize the significance of the ohmic heating caused by electron-neutral collisions during higher $\beta$ reconnection events below the middle chromosphere.
Figure~\ref{fig_2} also shows that viscous heating (Q$_{visc}$), has the lowest contribution to plasma heating for all values of $\beta_0$ in a reconnection process in the photosphere and lower chromosphere.

In the middle chromospheric (Z = 1000 km and 1400 km) cases, the variation of the heating terms with $\beta_0$ is different, especially in the higher $\beta$ domain.
Unlike magnetic reconnection below the middle chromosphere, the average values of all heating terms from Z = 1000 km - 1400 km (Figure~\ref{fig_2}(d) and \ref{fig_2}(e)) show almost similar decreasing trend with $\beta_0$.
Thus, we do not observe an enhancement of the joule heating (Q$_{en}$) or any other heating mechanism in any plasma-$\beta$ domain.
Therefore, Q$_{comp}$ is always the main heating mechanism to heat the plasma during the magnetic reconnection process when the reconnection region is in the middle chromosphere.
At Z = 1400 km, the amplitude of Q$_{vis-average}$ is still smaller than Q$_{comp-average}$, but it is larger than Q$_{ei-average}$ and Q$_{Amp-average}$ when $\beta_{0} \leq 0.5$ and is almost comparable to these heating terms when $\beta_{0} > 0.5$.
This means that the viscous heating mechanism is becoming important as the reconnection region moves up into the upper chromosphere. At the same time, Q$_{en}$ becomes the least important heating term above the middle chromosphere.


Simulations at the top of the chromosphere (Z = 2000 km) in Figure~\ref{fig_2}(f) show a unique evolution of energy densities with the initial plasma-$\beta$ compared to the other cases described above.
The compression heating (Q$_{comp-average}$) and viscous heating (Q$_{vis-average}$), generally show a decreasing pattern with $\beta_0$, and the decay gets faster when $\beta_0$ is larger.
The ambipolar heating (Q$_{Amp-average}$) and ohmic heating owing to electron-neutral collisions, Q$_{en-average}$ (green and red lines in the Figure~\ref{fig_2}(f), respectively) increase with $\beta_0$ before $\beta_0 \leq 0.5$ and then they decrease with $\beta_0$ like the other heating terms.
At such a high altitude in the upper chromosphere, we notice that the maximum temperature in the reconnection region exceeds 20,000 K in the cases with $\beta_0 < 0.5$, which causes strong ionization in the reconnection region, and the heating by ambipolar diffusion (Q$_{Amp-average}$) is not efficient.
However, viscous heating cannot be overlooked as its amplitude is not much lower than the compression heating term, even in low $\beta_0$ cases.
For magnetic reconnection with weaker magnetic fields ($\beta_0 > 0.5$), Q$_{comp-average}$, Q$_{vis-average}$, and Q$_{Amp-average}$ are of the same order of magnitude, suggesting that the heating caused by compression, viscosity, and ambipolar diffusion is equally important. The total value of Q$_{vis-average}$ and Q$_{Amp-average}$ exceeds Q$_{comp-average}$ in such a high $\beta_0$ range.

\begin{figure}
\centering
\begin{minipage}{0.494\textwidth}
\includegraphics[width=1.0\textwidth]{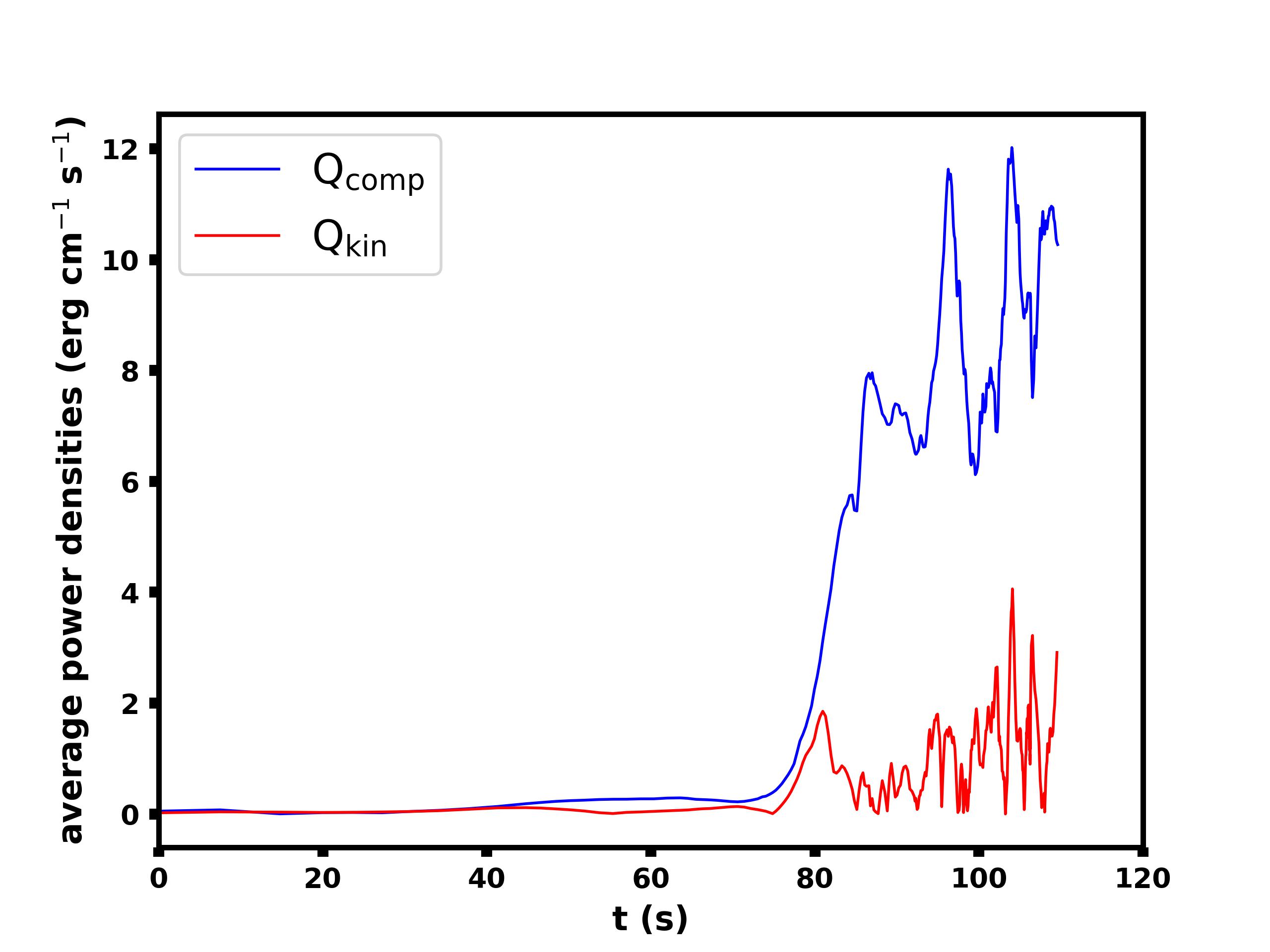}
\put(-160,175){\textbf{(a) Z = 800 km}}
\end{minipage}
\begin{minipage}{0.494\textwidth}
\includegraphics[width=1.0\textwidth]{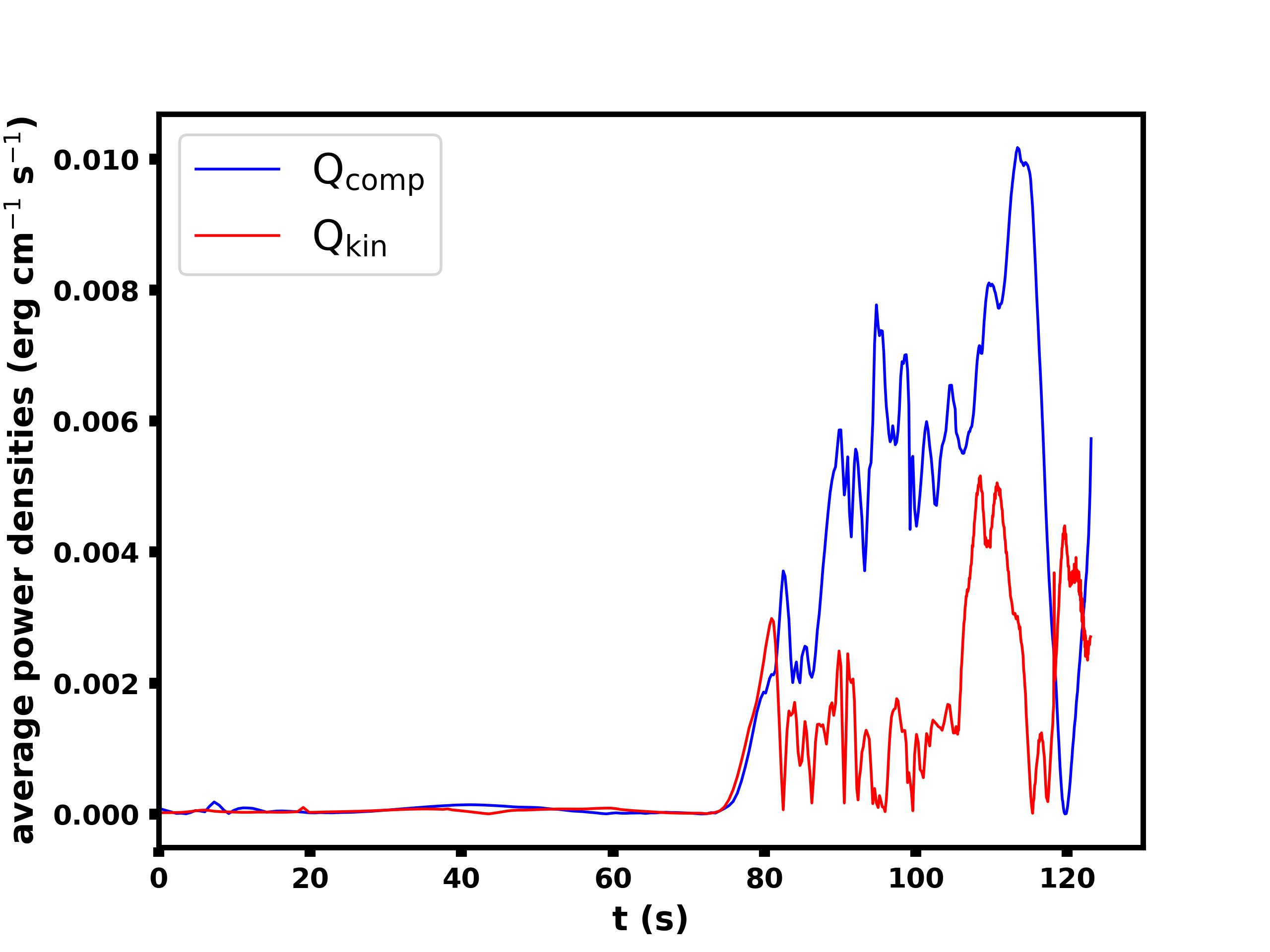}
\put(-160,175){\textbf{(b) Z = 2000 km}}
\end{minipage}
\caption{The temporal evolution of the compressional heating (Q$_{comp}$) and kinetic (Q$_{kin}$) power densities at (a) Z = 800 km and (b) Z = 2000 km with $\beta_0 = 0.5$.}
\label{fig_6}
\end{figure}

The results presented in Figure~\ref{fig_2} showed that in lower $\beta_0$ or stronger initial magnetic field domains, compression heating is the main heating mechanism during the turbulent magnetic reconnection process mediated with plasmoids at all altitudes in the lower solar atmosphere.
However, for the cases with weaker magnetic field and higher $\beta_0$, joule heating caused by $\eta_{en}$ will dominate to heat plasma in a magnetic reconnection process below the middle chromosphere, and the heating caused by ambipolar diffusion and viscosity is equally significant as that due to compression effect in the upper chromosphere.

The evolution of the average diffusion coefficients due to electron-ion collision ($\eta_{ei}$), electron-neutral collision ($\eta_{en}$), and ambipolar diffusion ($\eta_{Amp} = \eta_{AD} B^{2}/\mu_{0}$) versus $\beta_0$ are shown in Figure~\ref{fig_3}. 
The $\eta_{Amp-average}$ decreases with the initial plasma-$\beta$ at all different altitudes except at the top of the chromosphere (Z = 2000 km). 
An increase of $\eta_{ei-average}$ with $\beta_0$ is seen at all heights in the lower solar atmosphere, while $\eta_{en-average}$ first increases and then decreases with $\beta_0$.
$\eta_{ei}$ (inversely proportional to temperature, Eq.~\ref{eq:13}) and $\eta_{en}$ (depending on plasma ionization) are small in lower $\beta_0$ (stronger magnetic field) reconnection cases due to the significant plasma heating and strong ionization.
It is important to note that the average value of ambipolar diffusion ($\eta_{Amp-average}$) is higher at low $\beta_0$ and decreases when $\beta_0$ increases. 
This is due to the $B^{2}$ term in the calculation of ambipolar diffusion. 
$B^{2}$ is contributing more than $\eta_{AD}$ (depending on plasma ionization) with the modification of the magnetic field strength ($\beta_0$).
The average value of $\eta_{Amp}$ is higher than the average values of $\eta_{en}$ and $\eta_{ei}$ when $\beta_0 \sim 0.05$ at all altitudes below the top of the chromosphere. 
However, the role of the other diffusivities ($\eta_{en}$ and $\eta_{ei}$) becomes important as $\beta_0$ increases.
One should also note that the high value of $\eta_{Amp-averege}$ in the low $\beta_0$ case is contributed by the low temperature plasma in the reconnection region with non-uniform temperature distributions. 
The decreasing trend of $\eta_{Amp-averege}$ with $\beta_0$ below the top of the chromosphere (Figures~\ref{fig_3}(a)-\ref{fig_3}(e)) is consistent with the evolution of Q$_{Amp-averege}$ (Figures~\ref{fig_2}(a)-\ref{fig_2}(e)).
At the top of the chromosphere, $\eta_{Amp-averege}$ and $\eta_{en-averege}$ (Figure~\ref{fig_3}(f)) exhibit two different evolution stages, similar to those of Q$_{Amp-averege}$ and Q$_{en-averege}$ shown in Figure~\ref{fig_2}(f).

Figure~\ref{fig_4} presents the relation between Q$_{comp-average}$ and $\beta_0$ in six different layers, ranging from the photosphere to the top of the chromosphere.
The red stars indicate the values of Q$_{comp-average}$ calculated from simulations with different $\beta_0$, whereas the black dotted lines represent the power law fits as a reference.
A definite decay pattern for Q$_{comp-average}$ is found, and Q$_{comp-average}$ decreases as a power of $\beta_0$ in all cases at different altitudes.
As shown in Figure~\ref{fig_4}, the power law exponents vary from -1.90 at the photosphere to -1.29 at the top of the chromosphere.

As shown in Figure~\ref{fig_2}, the trends of Q$_{ei-average}$, Q$_{vis-average}$, and Q$_{Amp-average}$ exhibit a similar decay pattern with $\beta_0$ as Q$_{comp-average}$ in each case, except for the upper chromosphere at Z = 2000 km.
Q$_{Amp-average}$ increases with $\beta_0$ in the small $\beta_0$ rage in the upper chromosphere case.
The evolution of Q$_{en-average}$ as a function of $\beta_0$ is more complex.
In the photosphere and lower chromosphere, as $\beta_0$ increases in the range $0.04 < \beta_0 < 20.0$, Q$_{en-average}$ shows a trend of decreasing first, then increasing and finally decreasing again.
In the middle chromosphere, Q$_{en-average}$ always decreases as $\beta_0$ increases, and the decay trend closely resembles a single power law function.
In the upper chromosphere, the trend of Q$_{en-average}$ with variations in $\beta_0$ is similar to the trend of Q$_{Amp-average}$ as $\beta_0$ changes, Q$_{en-average}$ increases for $\beta_0 <0.5$ and decreases for $\beta_0 > 0.5$.
Since the joule heating contributed by $\eta_{en}$ is calculated as Q$_{en} = \eta_{en} J^2$, the value of Q$_{en}$ is determined by both the diffusion coefficients $\eta_{en}$ and the current density $J$.
A stronger magnetic field (lower $\beta_0$) will result in a larger current density $J$, but it causes a smaller diffusion coefficient $\eta_{en}$ (more neutrals become ionized) at the same time (Figure~\ref{fig_3}).
Therefore, when $\beta_0$ decreases, decreasing $\eta_{en}$ and increasing $J$ cause Q$_{en}$ to change in a very complex way with respect to $\beta_0$.

The evolution of the average thermal power density (Q$_{th-average}$) and kinetic power density (Q$_{kin-average}$) with initial plasma-$\beta$ at six different
altitudes are displayed in Figure~\ref{fig_5}. 
In the lower $\beta_0$ ($\beta_0 = 0.05$) cases, the average value of the kinetic power density is slightly higher than the average thermal power density when the reconnection region is below the middle chromosphere (Figures~\ref{fig_5}(a)-\ref{fig_5}(d)). 
When $\beta_0 > 0.1$, Q$_{th-average}$ is larger than Q$_{kin-average}$, which indicates that the generated thermal energy is larger than the kinetic energy during a reconnection process for higher plasma-$\beta$ case.  
Both (Q$_{th-average}$) and (Q$_{kin-average}$) decrease with $\beta_0$. 
The Q$_{kin-average}$ decays slightly faster than the Q$_{th-average}$, and the difference in their amplitudes increases as $\beta_0$ increases.

At Z = 2000 km (top of the chromosphere), the decay trend of the average thermal power density versus the initial plasma-$\beta$ is similar to that of the other altitudes, but the trend of the average kinetic power density is different. 
The simulation with $\beta_0 = 0.01$ is not well saturated and terminates earlier, which is why Q$_{kin-average}$ is significantly lower than Q$_{th-average}$ at $\beta_0 = 0.01$. The value of Q$_{kin-average}$ at $\beta_0 = 0.01$ is even lower than the average value at $\beta_0 = 0.05$ (Figure~\ref{fig_5}(f)).
It is possible that Q$_{kin-average}$ might increase further and reach somewhere close to Q$_{th-average}$ if the simulation runs longer.
Q$_{kin-average}$ then exhibits a similar decline pattern as observed at other altitudes (Figures~\ref{fig_5}(a)-\ref{fig_5}(e)) when $\beta_0 > 0.05$.  
These findings demonstrate the significance of the generated kinetic and thermal energies during the reconnection process, with kinetic energy dominating in low-$\beta$ plasma and thermal energy in high-$\beta$ plasma. 

The temporal evolution of compression heating (Q$_{comp}$) and kinetic energy (Q$_{kin}$) for two cases at different altitudes are shown in Figure~\ref{fig_6}.
In both simulation cases, each sharp increase in Q$_{comp}$ coincides with a sharp drop in Q$_{kin}$ during the fast reconnection stage with plasmoid instability, indicating that part of the kinetic energy resulted from magnetic reconnection is further converted to thermal energy by the compression process due to the turbulent flows in the reconnection region.


\section{Summary and discussion}
\label{sec_IV}
In this work, we have investigated the energy conversion mechanism during the magnetic reconnection process in the partially ionized solar plasma.
The reconnection current sheet with varying heights in the photosphere and chromosphere are considered to be parallel to the surface of the Sun.
The initial plasmas parameters (T$_0$, $\rho_0$) at different altitude are set according to the solar atmosphere model~\citep{avrett2008models}. At a particular height, we modify the strength of the magnetic field for each case with different initial plasma-$\beta$.
The results presented here explore the influence of the initial plasma-$\beta$ on various heating mechanisms that contribute to plasma heating in the reconnection process.
A critical plasma-$\beta$ ($\beta_{0-critical}$) is identified, where, in addition to compression heating, the joule heating contributed by electron-neutral collisions, ambipolar heating, and viscous heating also play an important role in the reconnection process at different altitudes.
Our main conclusions are as follows: 
\begin{enumerate}
    \item The lower plasma-$\beta$ will result in the generations of more plasmoids, higher frequency coalescence, stronger turbulence and local compression heating in the reconnection region.  
    \item The compression heating is generally the primary heating mechanism in the reconnection region in the lower plasma-$\beta$ environments. However, the joule heating contributed by electron-neutral collisions dominates over the compression heating when the initial plasma-$\beta$ exceeds the critical value ($\sim 8$) in the photosphere and lower chromosphere. In the upper chromosphere, the viscous heating and ambipolar diffusion heating will become equally important as the compression heating for $\beta>0.5$.
    \item The average value of the compression heating in the reconnection region decreases with the initial plasma-$\beta$ as a power function Q$_{comp-average} \sim \beta_0^{-a}$, the power index $a$ varies from $1.9$ in the photosphere to $1.29$ in the upper chromosphere. The joule heating Q$_{ei-average}$, viscous heating Q$_{vis-average}$, and ambipolar diffusion heating Q$_{Amp-average}$ also show a similar decreasing pattern with $\beta_0$ below the upper chromosphere. However, Q$_{en-average}$ evolves in a complex fashion at Z = 400 - 800 km, initially decreasing, then increasing, and finally decaying with $\beta_0$. Q$_{en-average}$ and Q$_{Amp-average}$ both show an increasing trend when $\beta_0 < 0.5$ at the top of the chromosphere.
    \item The generated kinetic energy dominates over the thermal energy during a magnetic reconnection process with $\beta_0 < 0.1$, while thermal energy exceeds kinetic energy as $\beta_0$ increases, highlighting the dominance of kinetic and thermal energies generation in different $\beta$ domains.

\end{enumerate}

The advanced radiative MHD simulations have well exhibited the typical characteristics of small-scale reconnection events such as EBs and UV bursts in the low solar atmosphere~\citep{danilovic2017simulating,hansteen2017bombs,hansteen2019ellerman,cheng2024magnetic}. 
However, the heating mechanisms in the reconnection region of these events have not been deeply explored in these fancy simulations. 
The early 2D numerical study showed that Joule heating is the main mechanism to heat plasma in a higher $\beta$ reconnection process, but the fixed ionization degree caused an overestimation of joule heating~\citep{ni2016heating}. 
Then, the numerical study with a more realistic radiative cooling and ionization model~\citep{liu2023numerical} proved that the joule heating contributed by electron-neutral collision can play the primary role to heat plasmas before the plasmoid instability takes place in a higher $\beta$ reconnection process, but compression heating will dominate over joule heating after the plasmoids appear. 
The fast reconnection and heating mechanisms at different altitudes from the photosphere to the upper chromosphere with a particular initial plasma-$\beta$ ($\beta_0=1.33$) have been deeply studied in recent work~\citep{zafar2023high}. However, the complex low solar atmosphere is filled with plasmas having different density, temperature and strength of magnetic fields that cause the plasma-$\beta$ varying within a wide range ($\sim 10^{-3} - 10^{3}$)~\citep{leake2014ionized}. In this work, we have performed many simulations at different altitudes and with plasma-$\beta$ varying from $0.01$ to $15$. We obtained the decreasing power law relationship between the averaged values of different heating terms in the reconnection region and the initial plasma-$\beta$. The critical initial plasma-$\beta$, above which compression heating is not the dominant heating mechanism, is also derived.

A plasma-$\beta$ threshold, below which compression heating remains the dominant heating mechanism, has been identified in the previous MHD simulation study with fully ionized plasma~\citep{birn2012role}.
They found that the contribution of joule heating exceeds that of compression heating for $\beta \sim 5-10$, which is almost comparable to the $\beta_{0-critical}$ value proposed for the magnetic reconnection below the middle chromosphere in the present work.
However, we would like to emphasize that a more realistic temperature-dependent resistivity ($\eta = \eta_{ei} + \eta_{en}$) is utilized in our work.
Another important point to highlight is that a single X-point has been observed in the entire simulation domain in Birn et al.~\citep{birn2012role}, which might underestimate the generation of compression heating.
In our simulations, multiple X-points are observed because of the generation of plasmoids, resulting in considerable compression heating due to their interaction.

According to our numerical simulations, we can speculate the dominant heating mechanism in different types of reconnection events in the low solar atmosphere. 
In the active region, the strength of reconnection magnetic fields is normally greater than 100 G and the plasma-$\beta$ is lower than $8$. Therefore, the compression heating caused by turbulent flows in the reconnection region is most likely the dominant heating mechanism for the active region EBs. 
The temperature in the UV burst ($\gtrsim 20,000$ K)  is much higher than that in the EB, and the plasma-$\beta$ there should be even lower. 
As pointed out in the previous paper~\citep{ni2022plausibility}, the compression heating should be the dominant heating mechanism for these high temperature UV bursts. 
However, there is another kind of EBs which are generated in the quiet Sun region, and the temperature increase in these events is usually lower than that in active region EBs. 
Since the strength of magnetic field in the quiet Sun region is normally about tens of gauss, the joule heating contributed by electron-neutral collisions in the reconnection region possibly becomes the primary heating mechanism for these quiet Sun region EBs. 
For cold surges triggered by magnetic reconnection above the middle chromosphere, we should not ignore the ambipolar diffusion heating and viscous heating.

Our studies provide valuable insights into different heating mechanisms during the magnetic reconnection process in the partially ionized solar plasma. 
However, the ideal simulation setups of a horizontal current sheet with initial uniform density and temperature cannot fully represent the reconnection diffusion region of a real event. 
In reality, density and gravity stratification, as well as turbulence caused by convective motions, can alter the dynamics of a normally inclined current sheet. Future simulations with more realistic setups are required to verify our conclusions and speculations. 
How more realistic interactions among different fluid species~\citep{wargnier2023multifluid,wargnier2025time} affect the results is also worth investigating.

\section*{Acknowledgments}
This research is supported by the National Key R\&D Program of China Nos. 2022YFF0503800 (2022YFF0503804) and 2022YFF0503003 (2022YFF0503000); the NSFC Grants 12373060 and 11933009; the Strategic Priority Research Program of the Chinese Academy of Sciences with Grant No. XDB0560000; the Basic Research of Yunnan Province in China with Grant 202401AS070044; the Yunling Talent Project for the Youth; the Yunling Scholar Project of the Yunnan Province and the Yunnan Province ScientistWorkshop of Solar Physics; Yunnan Key Laboratory of Solar Physics and Space Science under the number 202205AG070009; The work has been carried out at the National Supercomputer Center in Tianjin, and the calculations are performed on the Tianhe new generation supercomputer. The data analysis has been performed on the Computational Solar Physics Laboratory of Yunnan Observatories.

\bibliography{sample631}{}
\bibliographystyle{aasjournal}



\end{document}